\documentclass[manuscript]{aastex}
\usepackage{cases}

\newcommand{\fpp}[2]{\frac{\partial #1}{\partial #2}}
\newcommand{\vctr}[1]{\mbox{\boldmath $#1$}}
\newcommand{\tani}[1]{\mathrm{#1}}

\slugcomment{Not to appear in Nonlearned J., 45.}

\shorttitle{Eruption by Emerging Flux}
\shortauthors{Kaneko \& Yokoyama}

\begin{document}

\title{Simulation Study of Solar Plasma Eruptions Caused  \\
    by Interactions between Emerging Flux and Coronal Arcade Fields}


\author{T. Kaneko and T. Yokoyama}
\affil{Department of Earth and Planetary Science, The University of Tokyo, 
    7-3-1 Hongo, Bunkyo-ku, Tokyo, 113-0033, Japan\\
    e-mail: kaneko@eps.s.u-tokyo.ac.jp}

\begin{abstract}
We investigate the triggering mechanisms of plasma eruptions 
in the solar atmosphere due to
 interactions between  emerging flux and  coronal arcade fields 
by using two-dimensional MHD simulations.
We perform parameter surveys with respect to  arcade field height, 
 magnetic field strength, and  emerging flux location. Our results show that two possible mechanisms 
exist, and which mechanism is dominant depends mostly on emerging flux location.  
One mechanism appears when the location of  emerging flux 
is close to the polarity inversion line (PIL) of an arcade field. 
This mechanism requires reconnection between the emerging 
flux and the arcade field, as pointed out by previous studies.  
The other mechanism appears when the location of emerging flux is around the edge of an arcade field.
This mechanism does not require reconnection between the emerging flux and 
the arcade field but does demand reconnection in the arcade field above the PIL. 
Furthermore, we found that the eruptive condition for 
this mechanism can be represented by a simple formula. 
\end{abstract}

\keywords{filament eruption, CME}

\section{INTRODUCTION}
Many kinds of eruptive phenomena, 
such as flares, filament eruptions and coronal mass ejections (CMEs), 
can be seen in the solar atmosphere. 
Eruptive flares and CMEs are often observed with filament eruptions
(\citealp{2003ESASP.535..403G}, \citealp{2004ApJ...614.1054J}, \citealp{2012ApJ...744..168L}). 
It is widely believed that these eruptive phenomena are triggered
by the same global magnetohydrodynamic (MHD) process by which free 
magnetic energy stored in the coronal magnetic field is released as a form 
of thermal and/or kinetic energy.  

Many models and theories about the triggers have been 
proposed from both observational and theoretical studies \citep{2011LRSP....8....1C},
however, the exact mechanism by which eruptions occur is still under debates.
The interaction between newly emerging flux and  coronal magnetic fields 
is one candidate. Many observational studies report that
 newly emerging flux exists near the filament channel before the occurrence
of a filament eruption (e.g. \citealp{1995JGR...100.3355F}, \citealp{1999ApJ...510L.157W}, 
\citealp{2008ChA&A..32...56X}, \citealp{2012SoPh..277..185P}). 
\cite{1995JGR...100.3355F} found in a statistical study that quiescent filament eruptions 
tended to occur  when the polarity of  newly emerging flux and the 
magnetic field of a filament channel were favorable for reconnection.
The observational statistical study of \cite{2008ChA&A..32...56X}   
confirmed the results of \cite{1995JGR...100.3355F}
and found that the location of emerging flux relative to the filament channel 
strongly affected the occurrence of eruptions.

\cite{2000ApJ...545..524C} performed a theoretical
study by two-dimensional MHD simulations.  
They used the quadrupole magnetic structure, 
which consisted of a detached flux rope and an overlying magnetic arcade 
as an initial condition, and modeled emerging flux by changing 
the magnetic field at the bottom boundary.  
As a result, the system entered a non-equilibrium state and the flux rope was driven
upward by the reconnection beneath it. 

\citet{Lin2001JGR} analytically investigated the detailed conditions for 
the eruptions by interaction between the emerging flux and
the pre-existing coronal magnetic field which contains a flux rope. 
Though the analytic solutions of the equilibrium curves allow us to know 
whether the conditions are eruptive or not, 
the nature of eruptions in some cases are not clear. 
In order to get the physical understanding and to validate the eruptive conditions,
numerical simulations must be performed.

\cite{Notoya2007ASPC} demonstrated a theoretical model of a filament eruption 
associated with  emerging flux by using three-dimensional MHD simulations with gravity.
A flux tube buried in the convection zone entered  the corona by magnetic buoyancy in a self-consistent manner.
The emerging flux deformed the arcade field, which caused reconnection, and resulted in 
the formation and eruption of a flux rope. \cite{2012ApJ...760...31K}
showed that  emerging flux introduced into the PIL can destabilize 
sheared magnetic arcade fields in the corona and cause plasma eruptions
through the use of three-dimensional zero-$\beta $ MHD simulations. 
One of the differences between \cite{Notoya2007ASPC} and 
\cite{2012ApJ...760...31K} was the location of emerging flux; 
it was apart from the PIL in the former, while it was on the PIL in the latter.
 
In this paper we aim to determine the conditions under which eruptions occur 
due to the interaction between newly emerging flux and coronal arcade fields 
and express them by using observable variables such as the
magnetic flux, the location of flux emergence and the shear angle of coronal arcade fields. 
For this purpose, We perform MHD simulations, based on a model similar to that of 
\cite{Notoya2007ASPC}.
The parameters varied in our study are 
the magnetic scale height which relates to the shear angle of the arcade field,
the magnetic field strength and the location of the emerging flux.
We show  the simulation model in \S 2 and the results in \S 3.
In \S 4, we discuss the results  drawn from the simulations,
and compare  them  with those of other observational and theoretical 
studies.
 
\section{SIMULATION MODEL}

\subsection{Basic Equations}
We solve nonlinear, time-dependent,
resistive, compressible MHD equations 
in a rectangular computational box. 
The simulations are 2.5-dimensional, i.e., 
all three components in velocity and magnetic fields are 
taken into account in a two-dimensional $x,y$-domain.
The gas is assumed to be ideal with 
the ratio of specific heats $\gamma=5/3$ . 
Thermal conduction, radiative cooling and viscosity are absent in our simulations.
The equations are
\begin{equation}
\fpp{\rho }{t}+\nabla \cdot \left(\rho \vctr{v}\right)=0,
\end{equation}
\begin{equation}
\fpp{\left(\rho \vctr{v}\right)}{t}+\nabla \cdot \left( \rho \vctr{v}\vctr{v} 
+p\vctr{I}-\frac{\vctr{B}\vctr{B}}{4\pi }+\frac{B^2}{8\pi
}\vctr{I} \right) -\rho \vctr{g}=0,
\end{equation}
\begin{equation}
\frac{\partial }{\partial t}\left( \frac{p}{\gamma -1}+\frac{1}{2}\rho 
\vctr{v}^{2} +\frac{B^2}{8\pi }
  \right)
+\nabla \cdot \left[ \left( \frac{\gamma }{\gamma -1}p +\frac{1}{2}\rho 
\vctr{v}^{2}\right)\vctr{v}+\frac{c}{4\pi }\vctr{E} \times \vctr{B}
\right]-\rho \vctr{g}\cdot \vctr{v}=0,
\end{equation}
\begin{equation}
\fpp{\vctr{B} }{t}=-c\nabla \times \vctr{E},
\end{equation}
\begin{equation}
\vctr{E}=-\frac{1}{c}\vctr{v}\times \vctr{B}
 +\frac{4\pi \eta }{c^{2}} \vctr{J},
\end{equation}
\begin{equation}
\vctr{J}=\frac{c}{4\pi }\nabla \times \vctr{B},
\end{equation}
where $\vctr{I}$ is a unit tensor,  $\eta $ is the magnetic diffusion rate, $\vctr{g}=(0,-g_\tani{pho})$ is the gravitational
acceleration, and $g_{\tani{pho}} > 0$ is constant. 
Temperature is derived from the equation of state:
\begin{equation}
T=\frac{m}{k_{B}}\frac{p}{\rho },
\end{equation}
where $m$ is the mean molecular mass ($m=m_{p}$ is assumed where $m_{p}$ 
is the proton mass), and $k_{B}$ is the Boltzmann's constant. 
The normalized units of length, velocity, and density are
$H_{\tani{pho}}=310\tani{km}$, $C_{s,\tani{pho}}=12\tani{km/s}$, and 
$\gamma \rho _{\tani{pho}}=2.8\times 10^{-7}\tani{g/cm^{3}}$, respectively, where the subscript `pho' means the typical value in the photosphere, and
$H_{\tani{pho}}$ and $C_{s,\tani{pho}}$ are scale height and speed of sound,
respectively.
The other normalized units are combinations of these units:
 time is $H_{\tani{pho}}/C_{s,\tani{pho}}=26\tani{s}$,
temperature is $(m/k_{B}\gamma )C_{s,\tani{pho}}^{2}=10^{4}\tani{K}$,
magnetic field is $(\rho _{\tani{pho}}/\gamma )^{1/2}C_{s,\tani{pho}}=490\tani{G}$, and
gravitational acceleration is 
$C_{s,\tani{pho}}^{2}/H_{\tani{pho}}=\gamma g_{\tani{pho}}
=4.5 \times 10^{4}\tani{cm\cdot s^{-2}}$.  

We use  anomalous resistivity (e.g. \citealp{1994ApJ...436L.197Y}) 
to trigger a Petschek-type reconnection:
\begin{equation}
\eta = \left\{ \begin{array}{ll}
0  & (v_{d}\leq v_{c}) \\
\eta _{0}\left(v_{d}/v_{c}-1\right)^{2} & (v_{d}\geq v_{c}),
\end{array} \right.
\label{resis}
\end{equation}
where $v_{d}=mJ/(e\rho )$ is drift velocity, $e$ is elementary charge,  
and the normalized values of the parameters 
are $\eta _{0}=0.01$ and $v_{c}=1000$. 
We  restrict the value of $\eta $  to  $\eta _{\tani{max}}=1.0$.

\subsection{Initial Conditions}
\begin{figure}[htbp]
\begin{center}
\includegraphics[bb=0 0 566 433,scale=0.5]{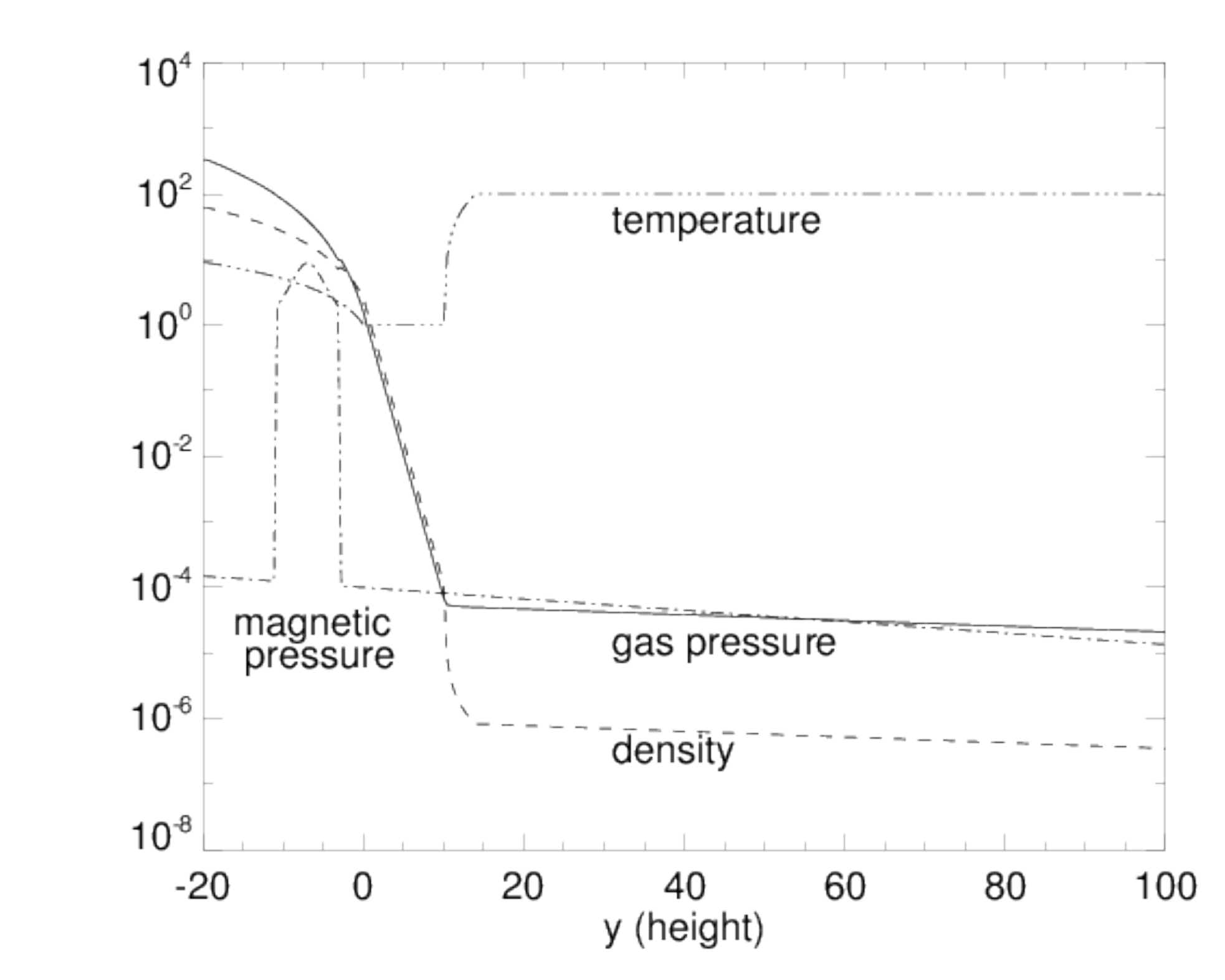}
\caption{Initial vertical distribution of each physical value. Magnetic pressure is plotted in 
the case of $B_{e}=15$.}
\label{init}
\end{center}
\end{figure}

The initial condition is the modeling of the solar atmosphere from the convection zone
to the corona. We follow the manner of modeling in the previous simulation 
studies \citep[e.g.][]{YokoyamaShibata1996PASJ}.
Fig.\ref{init} shows the initial vertical distribution of temperature, density, gas pressure, and 
magnetic pressure. 
The background gas consists of four layers:
\begin{numcases}{T(y)=}
T_{\tani{pho}}-\left( d\left|\frac{dT}{dy}\right|_{\tani{ad}} \right)y,~~(y\leq y_{\tani{pho}}) \\
T_{\tani{pho}},~~(y_{\tani{pho}}\leq y \leq y_{\tani{tr}}) \\
T_{\tani{pho}}+\left( T_{\tani{cor}}-T_{\tani{pho}}\right)\frac{y-y_{\tani{tr}}}{y_{\tani{cor}}-y_{\tani{tr}}},~~(y_{\tani{tr}}\leq y \leq y_{\tani{cor}})\\
T_{\tani{cor}},~~(y_{\tani{cor}} \leq y)
\end{numcases}
where $T_{\tani{pho}}=1~(10^{4}\tani{K})$, $T_{\tani{cor}}=100~(10^{6}\tani{K})$, 
$y_{\tani{pho}}=0$, 
$y_{\tani{tr}}=10~(3100\tani{km})$, and $y_{\tani{cor}}=14~(4340\tani{km})$. 
This  modeling of the solar atmosphere represents  
the subphotospheric convective zone, 
the photosphere/chromosphere, the transition layer and the corona, respectively. 
$|dT/dy| _{\tani{ad}}=(\gamma -1)/\gamma $ is the adiabatic gradient, and $d$ 
is a dimensionless constant which determines the stability of convection. 
Because $d$ is chosen to be 1.015, the vertical gradient of temperature
is slightly larger than the adiabatic temperature gradient,
causing the magnetic buoyancy instability to the flux tube.
The initial gas pressure is determined with the equation of gravitational stratification:
\begin{equation}
p(y)=p_{\tani{pho}}\exp \left[ -\int _{y_{\tani{pho}}} ^{y} 
\frac{mg_{\tani{pho}}}{k_{B}T(y^{\prime }
		 )}dy^{\prime }\right].
\end{equation}

We use a Gold-Hoyle type flux tube as a model of  emerging flux
\citep{2001ApJ...549..608M,2004ApJ...605..480M}.
In $(x-x_{e})^{2}+(y-y_{e})^{2}\leq r_{e}^{2}$, the magnetic field is given as 
\begin{eqnarray}
B_{x}&=&\frac{-B_{e}q(y-y_{e})}{1+q^{2}\left( [x-x_{e}]^{2}+[y-y_{e}]^{2}
			    \right)}, \label{goldx} \\
B_{y}&=&\frac{B_{e}q(x-x_{e})}{1+q^{2}\left( [x-x_{e}]^{2}+[y-y_{e}]^{2}
			    \right)}, \label{goldy} \\
B_{z}&=&\frac{B_{e}}{1+q^{2}\left( [x-x_{e}]^{2}+[y-y_{e}]^{2}
				    \right)}, \label{goldz}
\end{eqnarray}
where $(x_{e},y_{e})$ is the location of the axis of the flux tube,
$y_{e}=-7~(2170\tani{km})$, radius $r_{e}=4~(1240\tani{km})$, 
pitch number $q=0.5$, and the length of the one helical turn along the axis 
is $2\pi /q=4\pi~(3896\tani{km})$. 
In our simulations, $B_{e}$ takes the values of $11,13,15~\tani{and }~17~(5300\tani{G}-8300\tani{G})$, 
and $x_{e}$ is set in the range from $-40$ to $40$. 
This flux tube model satisfies the force-free condition inside itself, 
so the initial gas pressure in the flux tube is
\begin{equation}
p_{\tani{tube}}=p(y)-\frac{B_{e}^{2}}{8\pi }\frac{1}{1+q^{2}r_{e}^{2}}
\end{equation} 
for the pressure balance in a horizontal direction between the inside and the outside of 
the flux tube.
The temperature of the flux tube is uniform horizontally so 
that the density of the flux tube is lower than its outside, 
causing vertical force imbalance regarded as the magnetic buoyancy.

The periodic set of magnetic arcades as follows is superposed on the flux tube, 
\begin{eqnarray}
B_{x}&=&\left( \frac{2L}{\pi a}\right)B_{a}\sin \left( \frac{\pi }{2L}x
		\right)\exp \left[ -\frac{y}{a} \right], \label{ar1}\\
B_{y}&=&B_{a}\cos \left( \frac{\pi }{2L}x \right)
 \exp \left[ -\frac{y}{a} \right], \label{ar2}\\
B_{z}&=&-\sqrt{1-\left( \frac{2L}{\pi a} \right)^{2}}B_{a}\sin \left( 
		  \frac{\pi }{2L}x\right) \exp \left[ -\frac{y}{a}
					      \right], \label{ar3} 
\label{ar3}
\end{eqnarray}
where $B_{a}=0.05~(24\tani{G})$, $L=40~(1.2\times 10^{4}\tani{km})$, and
$a$ is set in the range from $2L/\pi =25.5$ (i.e., potential field) to 150. 
$L$ and $a$ represent the half width and the magnetic scale height 
of the arcade field, respectively.
In this paper, we define the  shear angle of the arcade field $\theta $ as
\begin{equation}
\theta =\tan ^{-1}\left(-\frac{B_{z}}{B_{x}} \right)
=\tan ^{-1}\left( \sqrt{\left(\frac{\pi a}{2L}\right)^{2}-1}\right).
\label{dtheta}
\end{equation}
For constant $L$, $\theta $ is increasing with $a$. 
Because the arcade field is also force-free, 
the initial condition is in a mechanical equilibrium. 

\begin{figure}[htbp]
  \begin{center}
    \includegraphics[bb=0 0 348 282,scale=0.4]{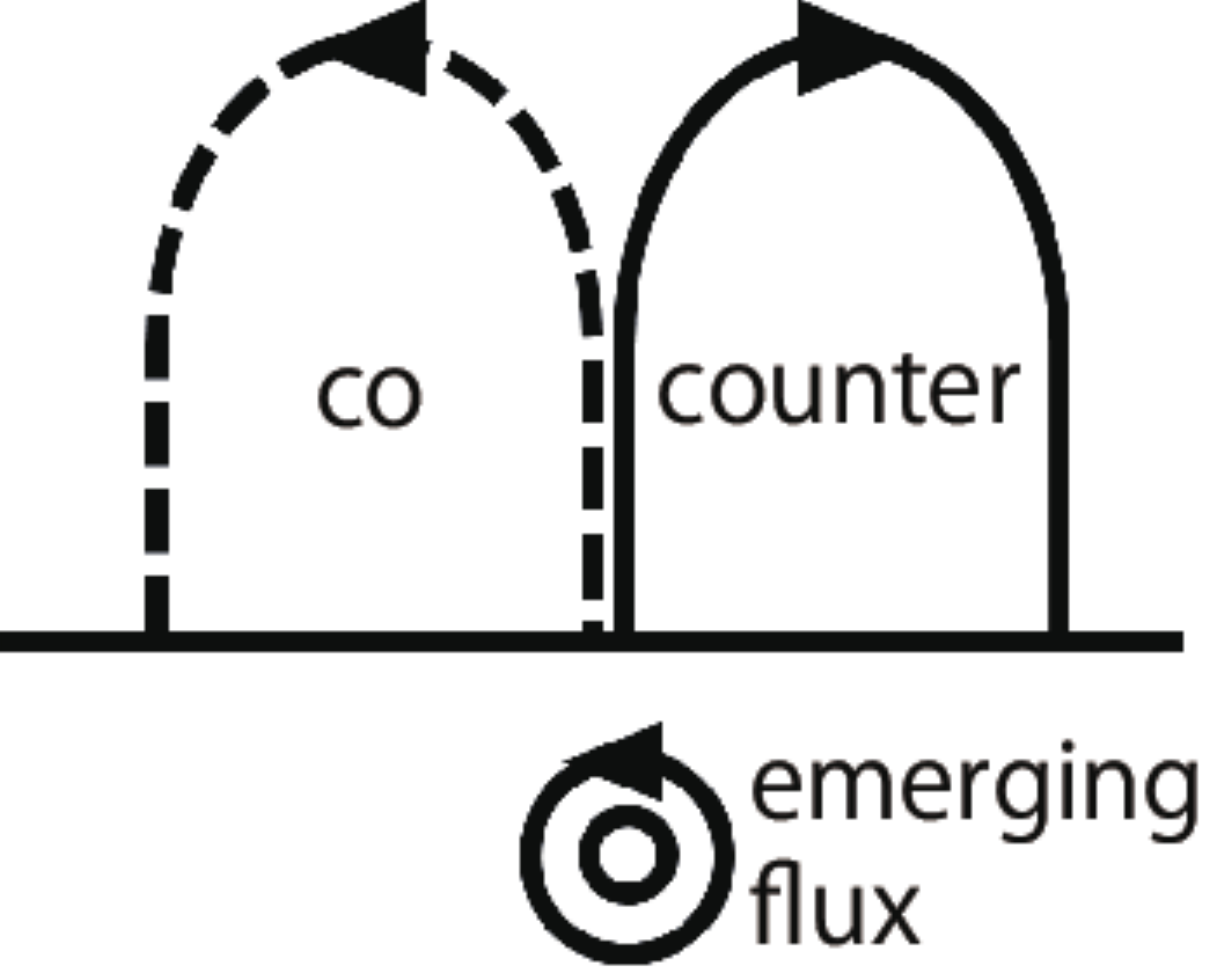}
    \caption{Positions of co-arranged and counter-arranged arcade fields. 
      The emerging flux is located at $x_{e}=0$ in this picture, and 
      the dashed arcade is co-arranged, and the solid arcade is counter-arranged.}
    \label{arc}
  \end{center}
\end{figure}

As for the relative arrangement between the emerging flux and the arcade fields,
we refer to the right arcade toward the flux tube in the initial condition as
`counter-arranged arcade'
because the polarity arrangement is in reverse order, 
and the left arcade as `co-arranged arcade' 
because the polarity arrangement is in the same order as emerging flux (Fig.\ref{arc}).

The horizontal grid resolution is uniform, with $\Delta x=0.4$. 
The vertical grid width is uniform, with $\Delta y=0.4 $ in $y<160$, 
and with  $\Delta y=6.0 $ in $y>300$.
The vertical grid width in $160<y<300$ gradually increases from 
$\Delta y=0.4$ to $6.0$. 
The simulation box is $-120<x<200$ and $-20<y<1204$. 
The calculation scheme is a modified Lax-Wendroff method
(Rubin \& Burstein,1967) with  artificial viscosity (Lapidus,1967).
A periodic boundary condition is applied to the left and right boundaries, and 
a free boundary condition is applied to the top and the bottom boundaries.

\section{RESULTS}
For the parameter survey, 
we performed the multiple simulations among which 
the field strength $B_{e}$, the location of the emerging flux $x_{e}$ and the
magnetic scale height of the arcade field $a$ varies.
The field strength of the emerging flux varies from $B_{e}=11$ to $17$.
The location of the emerging flux varies between $x_{e}=-40$ (at PIL of the co-arranged arcade)
and $x_{e}=40$ (at PIL of the counter-arranged arcade) at intervals of $10$. 
The magnetic scale height of the arcade field varies 
from $a=2L/\pi=25.5$ (for a potential field) 
to $150$ (when the shear angle is 80 degree).
In some cases, the plasmoids are formed and erupted by the interaction between 
the emerging flux and the arcade fields. We found that two types of eruptive mechanisms exist. 
Before showing the results of the parameter survey,
the typical examples of each eruptive mechanism are shown in the next subsection.

\subsection{Two mechanisms for formation and eruption of plasmoids}
Our simulations suggest two mechanisms for 
the formation and eruption of plasmoids: 
One is due to compression of the arcade field by the emerging flux 
(hereafter `CA-type' mechanism), and the other is due to 
reconnection coupling as a consequence of 
multipoint reconnections between the arcade field and the emerging flux 
(hereafter `RC-type' mechanism). The CA type eruptions can be seen 
when the location of the emerging flux is around $x_{e}=0$ (separator line
dividing the counter-arranged arcade and the co-arranged arcade).
The RC type eruptions can be seen when the location of the emerging flux 
is around $x_{e}=40$ (the PIL of the counter-arranged arcade).
In both mechanisms, the plasmoid is accelerated 
by  reconnection outflow. In spite of the distinct mechanisms, 
both the CA-type and the RC-type eruptions succeed in creating a hot cusp 
structure (\citealp{1992PASJ...44L..63T}), as proposed in the standard 
CSHKP model (\citealp{1964NASSP..50..451C}, \citealp{1966Natur.211..695S}, 
\citealp{1974SoPh...34..323H}, \citealp{1976SoPh...50...85K}). 

\begin{figure}
  \begin{center}
    \includegraphics[bb=0 0 909 907,scale=0.5]{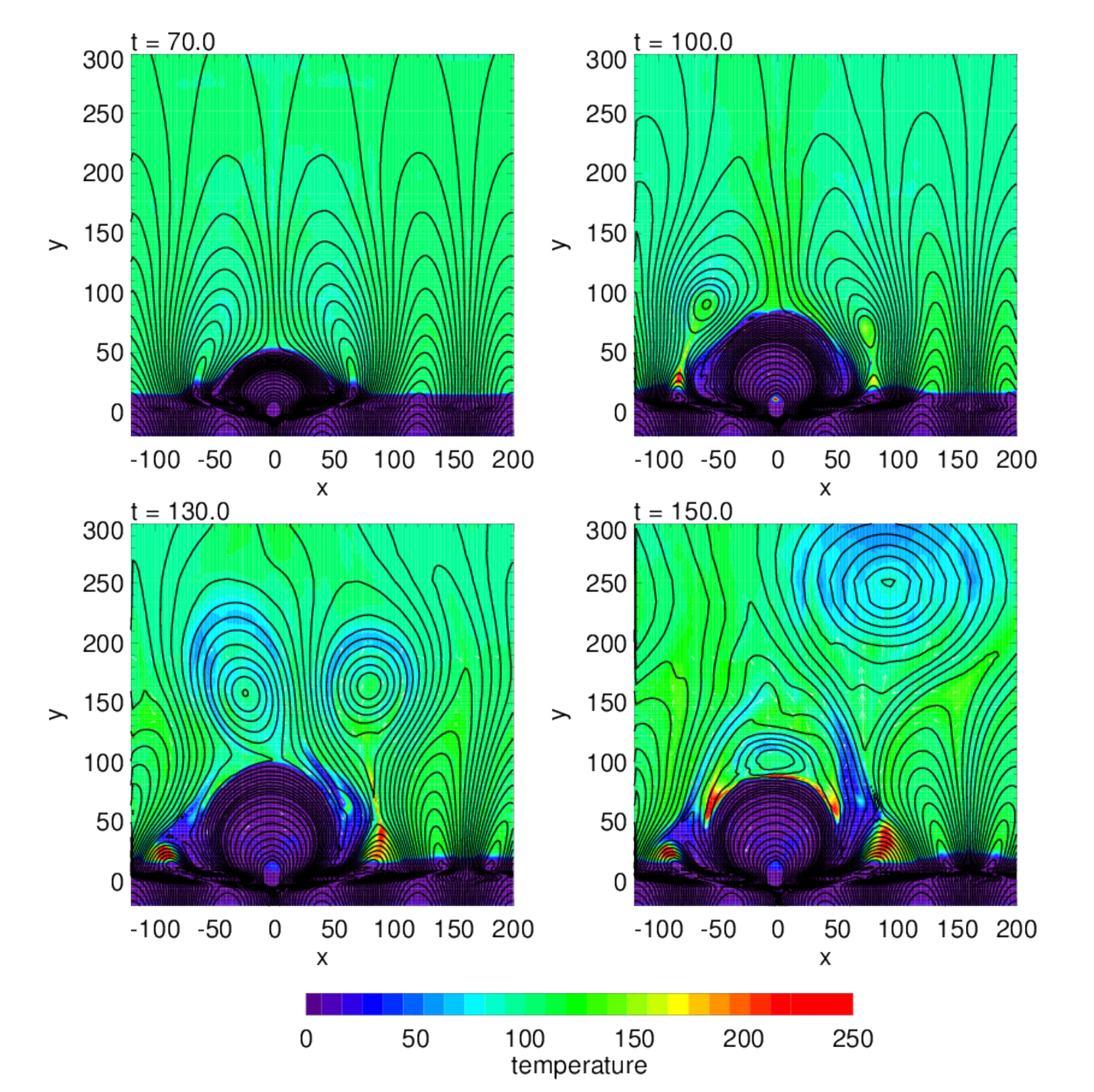}
    \caption{\small{Example of CA-type mechanism. 
        Time Evolution of $(B_{e},x_{e},a)=(15,0,100)$.
        Black lines represent magnetic field lines. Color 
        represents temperature. This figure is also available as mpeg animation
        in the electronic edition of the \textit{Astrophysical Journal}.
    } }
    \label{mecha1}
  \end{center}  
\end{figure}

\begin{figure}
  \begin{center}
    \includegraphics[bb=0 0 596 485,scale=0.7]{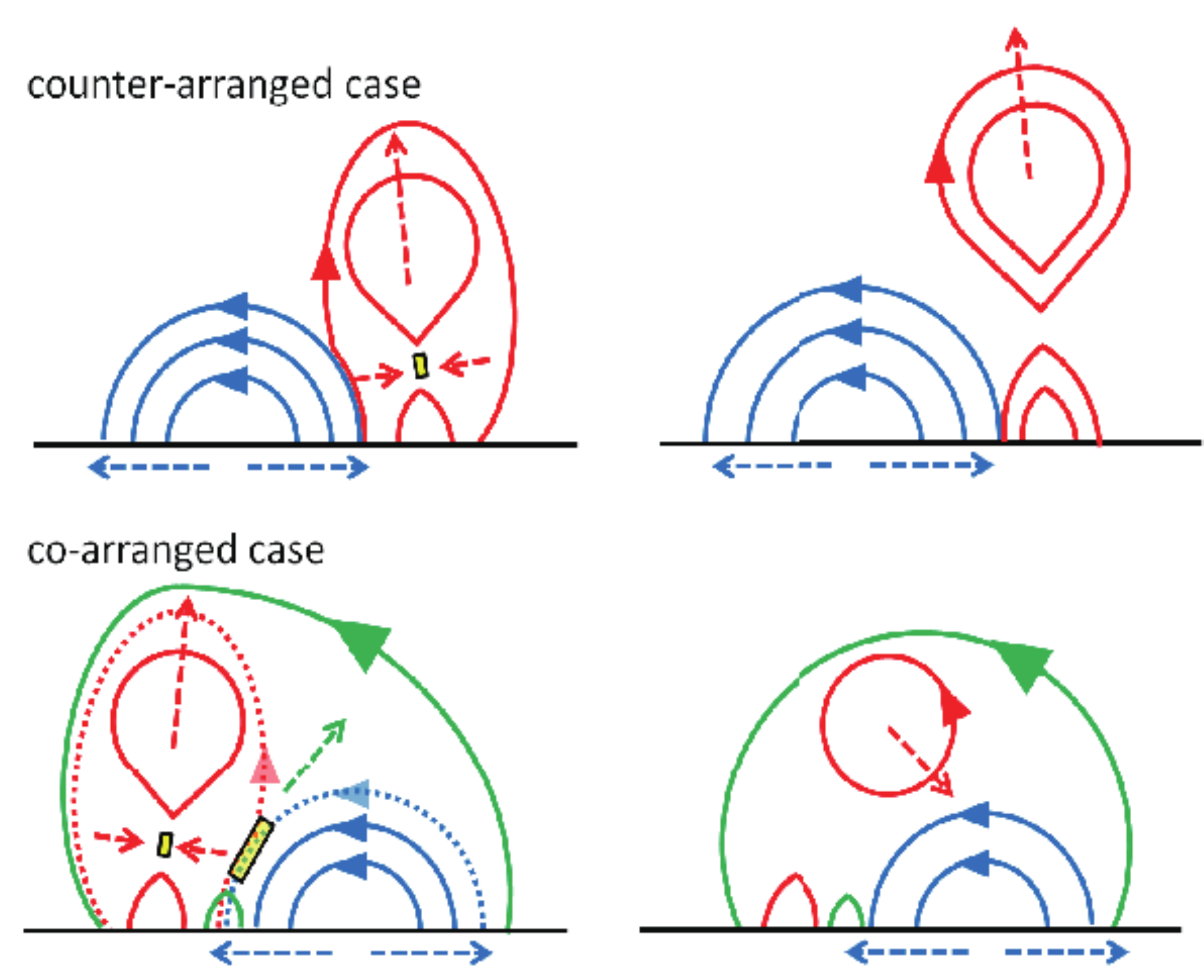}
    \caption{\small{Schematic pictures of the processes of the CA type eruption
        from the counter-arranged arcade and the confined eruption from 
        the co-arranged arcade. In the co-arranged case, 
        the reconnection between the field line of the arcade field (red dotted line) 
        and that of the emerging flux (blue dotted line) forms the  
        large overlying arcade fields (green solid line).
    } }
    \label{ca_mecha}
  \end{center}  
\end{figure}

Fig.\ref{mecha1} shows a time evolution of the CA-type mechanism in the case of 
$(B_{e},x_{e},a)=(15,0,100)$. The flux tube initially buried 
in the convection zone expands rapidly after entering 
the corona ($t=70$). 
The compression of the arcades by the expanding emerging flux makes
the current inside the arcades larger and thinner ($t=100$). 
As a result, reconnections occur inside both the counter-arranged and
co-arranged arcades above each PIL around $x=\pm 40$, 
leading to the formation of plasmoids.  
Both the plasmoids are initially accelerated by reconnection outflow ($t=130$).  
The plasmoid originating from the counter-arranged arcade continues to rise up 
even after reconnection stops ($t=150$). However, the plasmoid originating from
the co-arranged arcade is pulled by a downward magnetic tension force from the overlying
arcade field, resulting in the confined eruption. 
Since the field lines of the emerging flux and 
those of the co-arranged arcade are antiparallel, the reconnection is triggered in the
same manner as the standard jet model 
\citep{YokoyamaShibata1995Nature,YokoyamaShibata1996PASJ}. 
This reconnection transfers the magnetic flux from the right footpoints of the
co-arranged arcade to the right side of the emerging flux, 
forming the large overlying arcade (see also Fig.\ref{ca_mecha}). 
Even though the plasmoid is formed in the same manner as the counter-arranged arcade
case, the downward tension force of the overlying arcade prevents the plasmoid from
going upward. 
Consequently, the plasmoid turns to fall down after the temporal upward motion.

\begin{figure}
  \begin{center}
    \includegraphics[bb=0 0 624 481,scale=0.75]{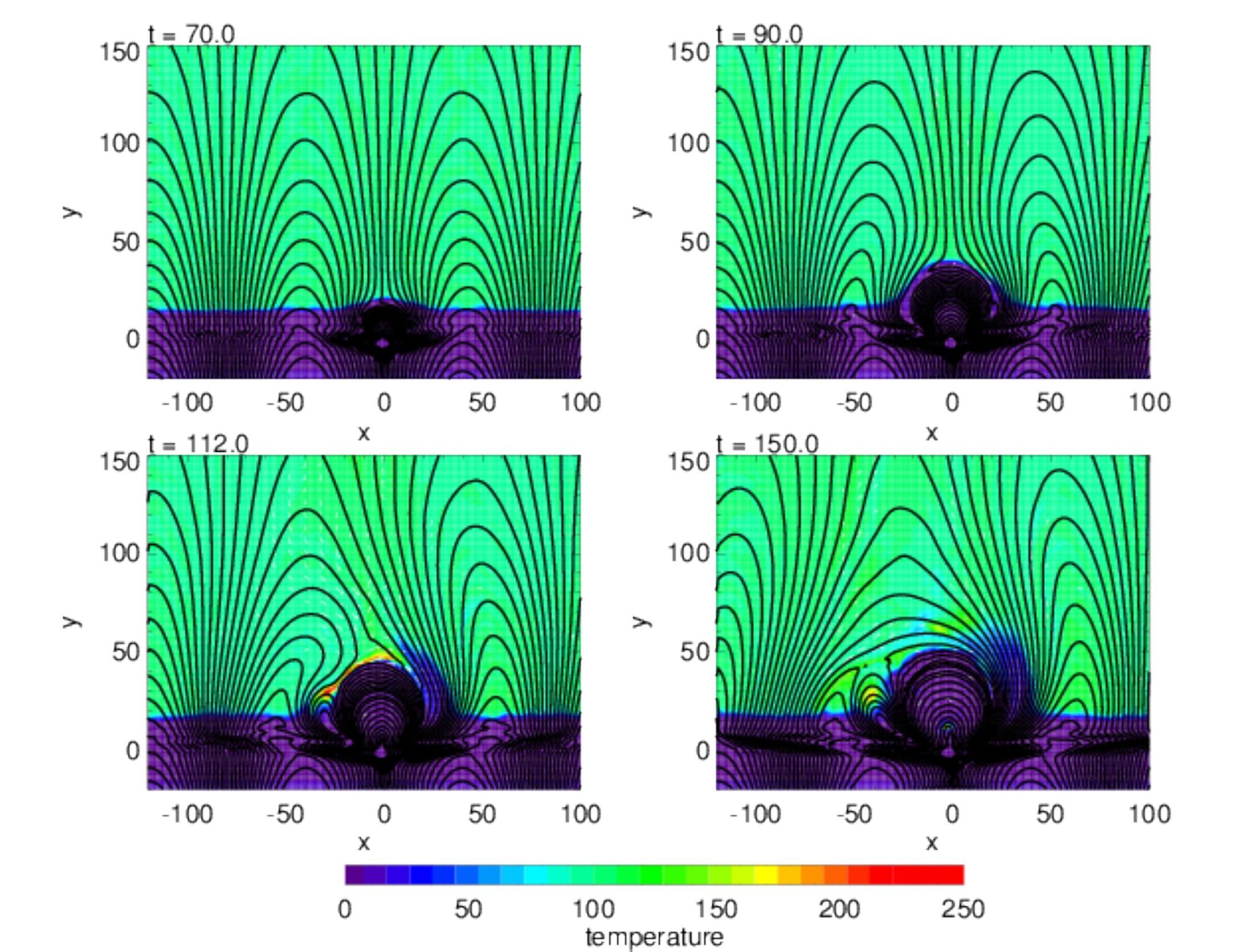}
    \caption{\small{Example of unformed case. 
        Time Evolution of $(B_{e},x_{e},a)=(11,0,100)$. Only the area around the 
        flux emergence is shown. Black lines represent magnetic field lines. 
        Color represents temperature.
    } }
  \end{center}  
  \label{unformed}
\end{figure}

Fig.\ref{unformed} shows a time evolution of $(B_{e},x_{e},a)=(11,0,100)$. 
This case is categorized to the unformed case. Compared with the erupted case such as  
$(B_{e},x_{e},a)=(15,0,100)$, the compression of the arcade is not sufficient
to trigger the reconnection inside the arcades because the field strength of the
emerging flux is small and its expansion in the corona is weaker.
Although weak reconnection is triggered 
between the co-arranged arcade and the emerging flux as well ($t=112$), 
the system rather evolves into a quasi-static state ($t=150$).

\begin{figure}
  \begin{center}
    \includegraphics[bb=0 0 909 907,scale=0.5]{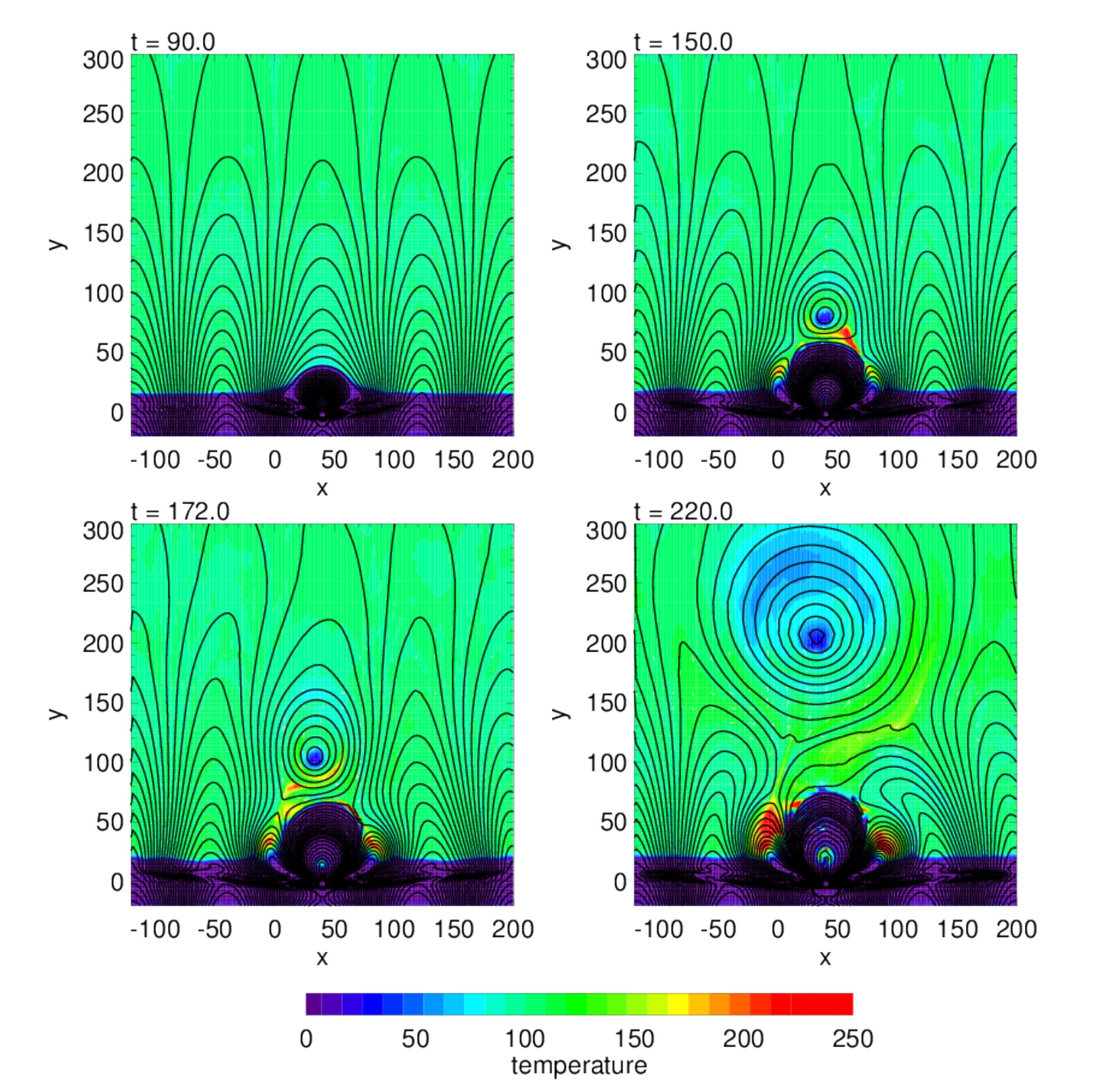}
    \caption{\small{Example of RC-type mechanism. 
        Time Evolution of $(B_{e},x_{e},a)=(11,40,100)$.
        Black lines represent magnetic field lines. Color
        represents temperature. This figure is also available as mpeg animation
        in the electronic edition of the \textit{Astrophysical Journal}.
    } }
    \label{mecha2}
  \end{center}  
\end{figure}

\begin{figure}
  \begin{center}
    \includegraphics[bb=0 0 425 188,scale=1.0]{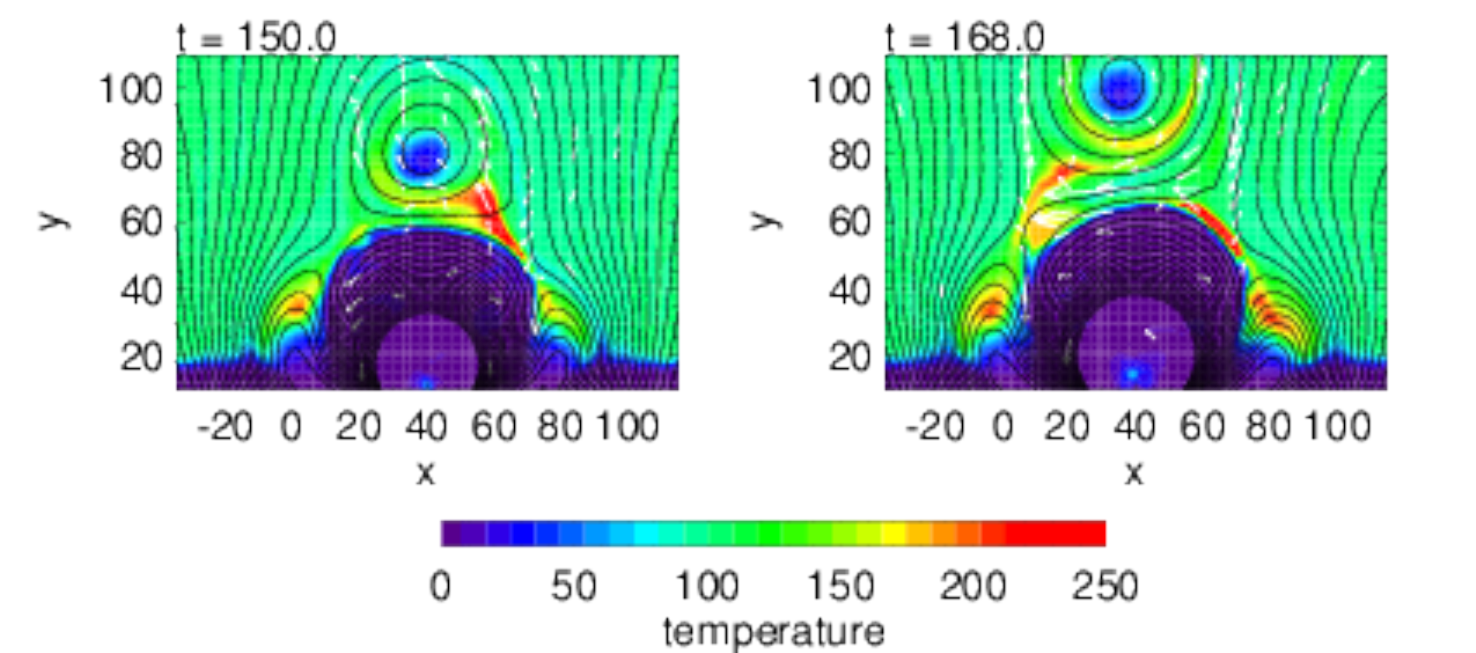}
    \caption{\small{Example of reconnection coupling
        in the case of $(B_{e},x_{e},a)=(11,40,100)$. White arrows represent velocity field.
        Colors represent temperature. Black lines are magnetic fields.
        The outflow from the right reconnection point 
        ($(x,y)=(70,55)$) passes through the boundary between 
        the arcade and the emerging flux
        and links to the inflow region of  the left reconnection point
        ($(x,y)=(5,60)$).
    } }
    \label{mecha2_bh}
  \end{center}  
\end{figure}

Fig.\ref{mecha2} shows a time evolution of the RC-type mechanism 
in the case of $(B_{e},x_{e},a)=(11,40,100)$.
Compared with the two cases shown above, the emerging flux is
located at the PIL of the counter-arranged arcade field.
In contrast to the CA-type, since the location of $x_{e}$ is at the PIL,
the emerging flux does not lead to a sideways compression  of the arcade field
and the formation of a vertical current sheet. 
In this case, a current sheet is formed at the interface between the  
emerging flux and the arcade field as these oppositely directed fields are pushing
against each other ($t=90$).
The reconnections are triggered at multiple points in this sheet
almost simultaneously, and a plasmoid grows ($t=150$). 
In the end, two reconnections play an important role for the eruptions. 
The hot outflow from a right-side 
reconnection point, which is slightly more intense compared with that from the
left-side reconnection point, 
passes through the boundary region between the arcade and the emerging flux, 
and links to the inflow region of the left-side reconnection point ($t=172$).
The left-side reconnection is driven by the outflow from the
right-side reconnection point, leading to a plasmoid eruption ($t=220$).
Fig.\ref{mecha2_bh} is the extended view of the reconnection coupling. 
The outflow from one reconnection point at $(x,y)=(70,55)$ reaches 
the other X-point at $(x,y)=(5,60)$ and promotes
the reconnection there, leading to the growth and eruption of a plasmoid.
Reconnection points in the current sheet appear in random positions as a 
result of a resistive instability in which the seed perturbations are imposed 
by numerical noises.           
This mechanism is similar to the breakout model \citep{Antiochos1999ApJ}
with respect that the reconnection between the overlying arcade field and
the expanding magnetic flux (which corresponds to the emerging flux in our simulations)
triggers the eruption. 
The critical difference is the number of
the reconnection points. The breakout model has one reconnection point (null point)
on the top of the expanding flux, while the RC-type mechanism has 
two reconnection points in the current sheet between the arcade 
field and the emerging (expanding) flux. This difference leads to    
the distinct eruptive process. In the breakout model, 
the expanding flux itself erupts, while, in the RC-type mechanism, 
the plasmoid formed by the two reconnections in the current sheet finally erupts.
Note that when the flux emergence is located at the PIL of the
co-arranged arcade ($x_{e}=-40$),
the reconnection between the emerging flux and the co-arranged arcade does not
occur because they have the same direction of the field lines.
The emerging flux just expands the arcade,
and ends up with a quasi-static state.

\subsection{Parameter Survey}
\begin{figure}
\begin{tabular}{c}
  \begin{minipage}{1.0\hsize}
    \begin{center}
      \includegraphics[bb=0 0 850 566,scale=0.4]{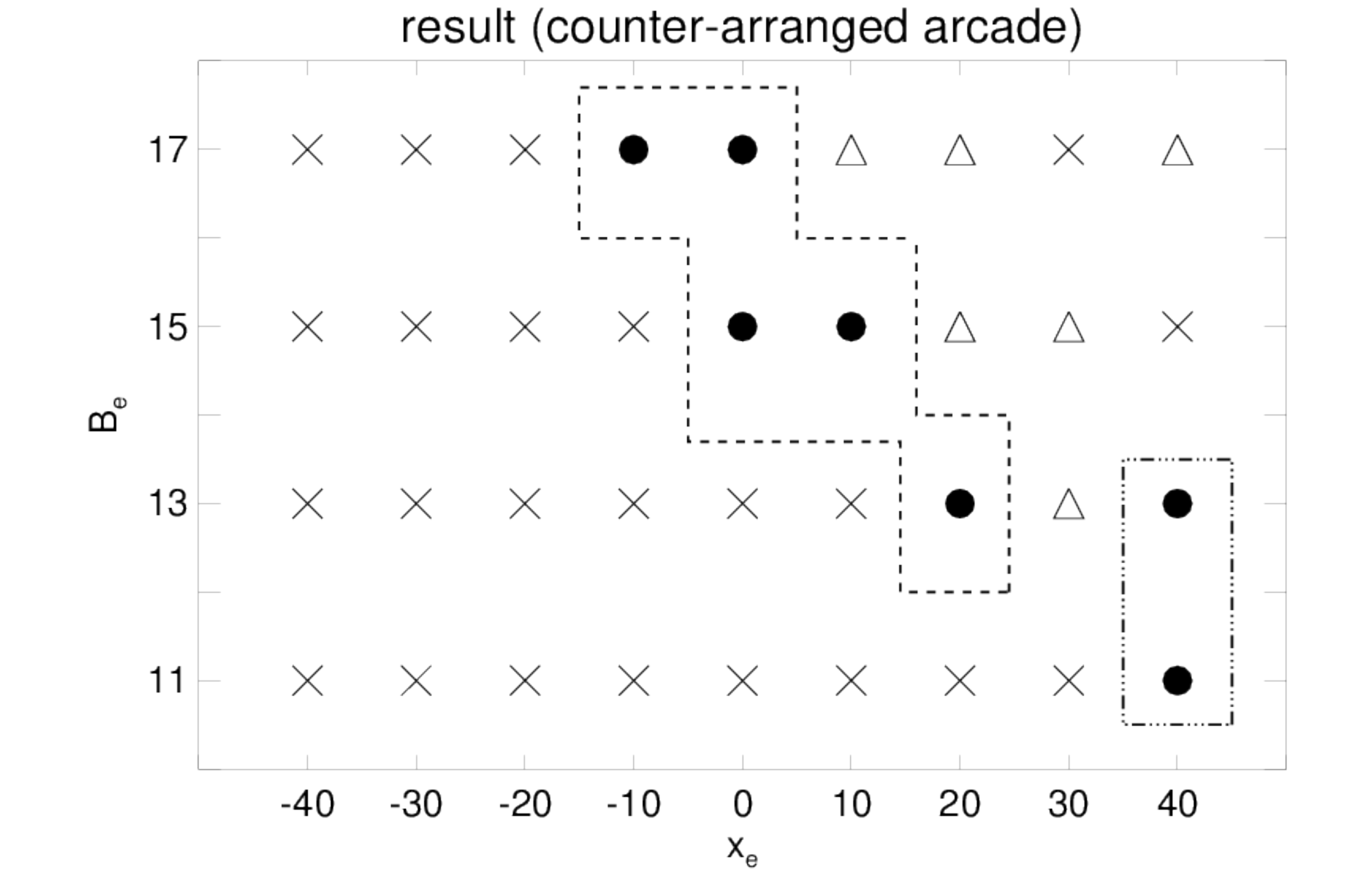}
    \end{center}
  \end{minipage}
  \\
  \\
  \begin{minipage}{1.0\hsize}
    \begin{center}
      \includegraphics[bb=0 0 850 566,scale=0.4]{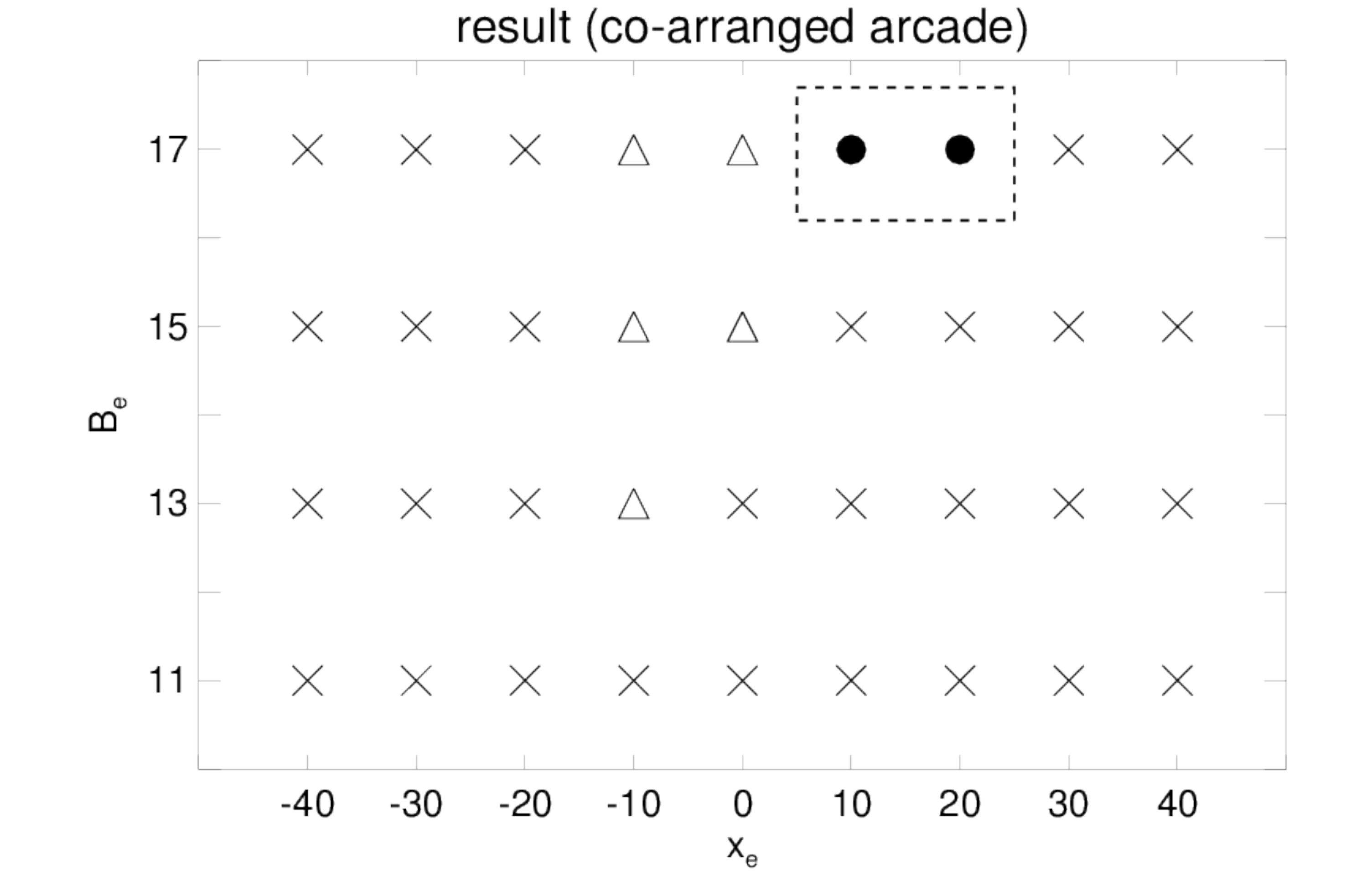}
    \end{center}
  \end{minipage}
  \end{tabular}
  \caption{\small{
      Results of parameter survey on field strength $B_{e}$ 
      and  location of  emerging flux $x_{e}$ for $a=100$. 
      The upper panel shows  results of the counter-arranged arcade
      and the lower panel shows results  of the co-arranged arcade. 
      Circles are `erupted', triangles are `formed', and crosses are  `unformed' cases. 
      Circles in dashed boxes represent  CA-type eruptions and those in dot-dashed boxes 
      represent  RC-type eruptions. }}
  \label{resr}  
\end{figure}

The result of each run is categorized to `erupted case', `formed case' or `unformed case'.
In the erupted cases, the formed plasmoids continuously go upward for the simulation time.
In the formed cases, the formed plasmoids turn to descend or stop after 
temporal ascending. The formed cases include the confined eruptions.
The unformed cases do not show the formation of any plasmoids.
The results of our parameter survey with respect to $B_{e}$ and $x_{e}$
are shown in Fig.\ref{resr}. The magnetic scale height of the arcade field is fixed
to be $a=100$. 
The filled circles represent erupted cases, 
the triangles represent the formed cases,
and the crosses represent the unformed cases.
CA-type eruptions occur when the
field strength of the emerging flux is sufficiently large, and its location 
is closer to the boundary between neighboring arcades.
However, when the flux emergence is too close to the PIL of an arcade, 
CA-type eruptions do not occur. 
RC-type eruptions occur only when the location of the
emerging flux is at the PIL of the counter-arranged arcade,
and when the field strength is sufficiently small.
There are several cases in which plasmoids form in both arcades, however,
in none of them both plasmoids erupt (see Section 4.4).

Fig.\ref{resa} shows the results of the parameter survey with respect 
to the magnetic scale height $a$. The cases of $(B_{e},x_{e})=(15,0)$ and
$(B_{e},x_{e})=(11,40)$ are investigated as the typical cases of the CA-type
eruptions and the RC-type eruptions, respectively.
Both the CA-type eruptions and the RC-type eruptions are more likely to
occur when $a$ is larger, i.e., when the shear angle of the arcade is larger.

\subsection{Ejection Speed of Plasmoids}
\begin{figure}
  \begin{center}
    \begin{tabular}{cc}
      \begin{minipage}{0.5\hsize}
        \includegraphics[bb=0 0 566 425,scale=0.41]{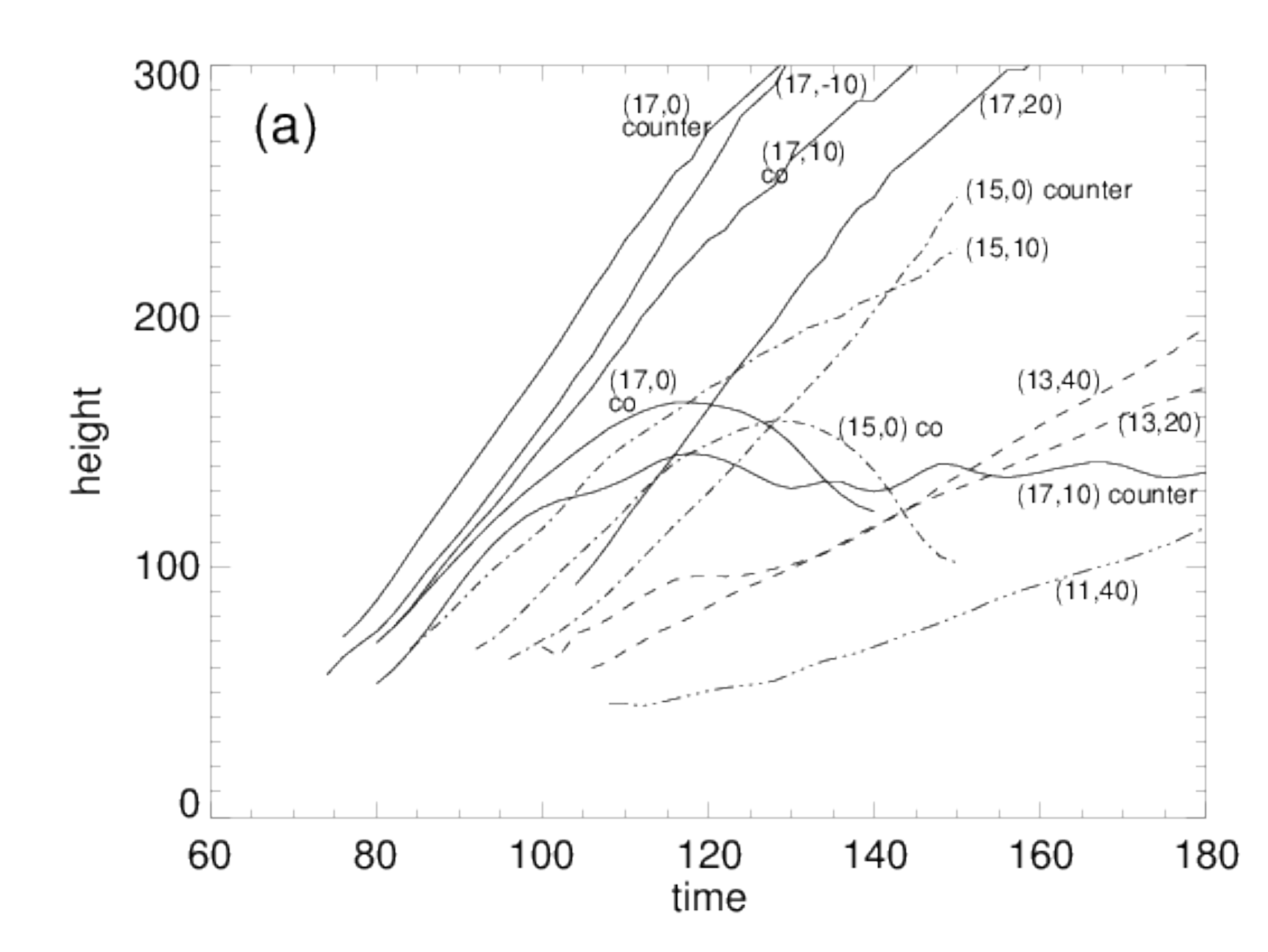}
      \end{minipage}
      \begin{minipage}{0.5\hsize}
        \includegraphics[bb=0 0 566 425,scale=0.41]{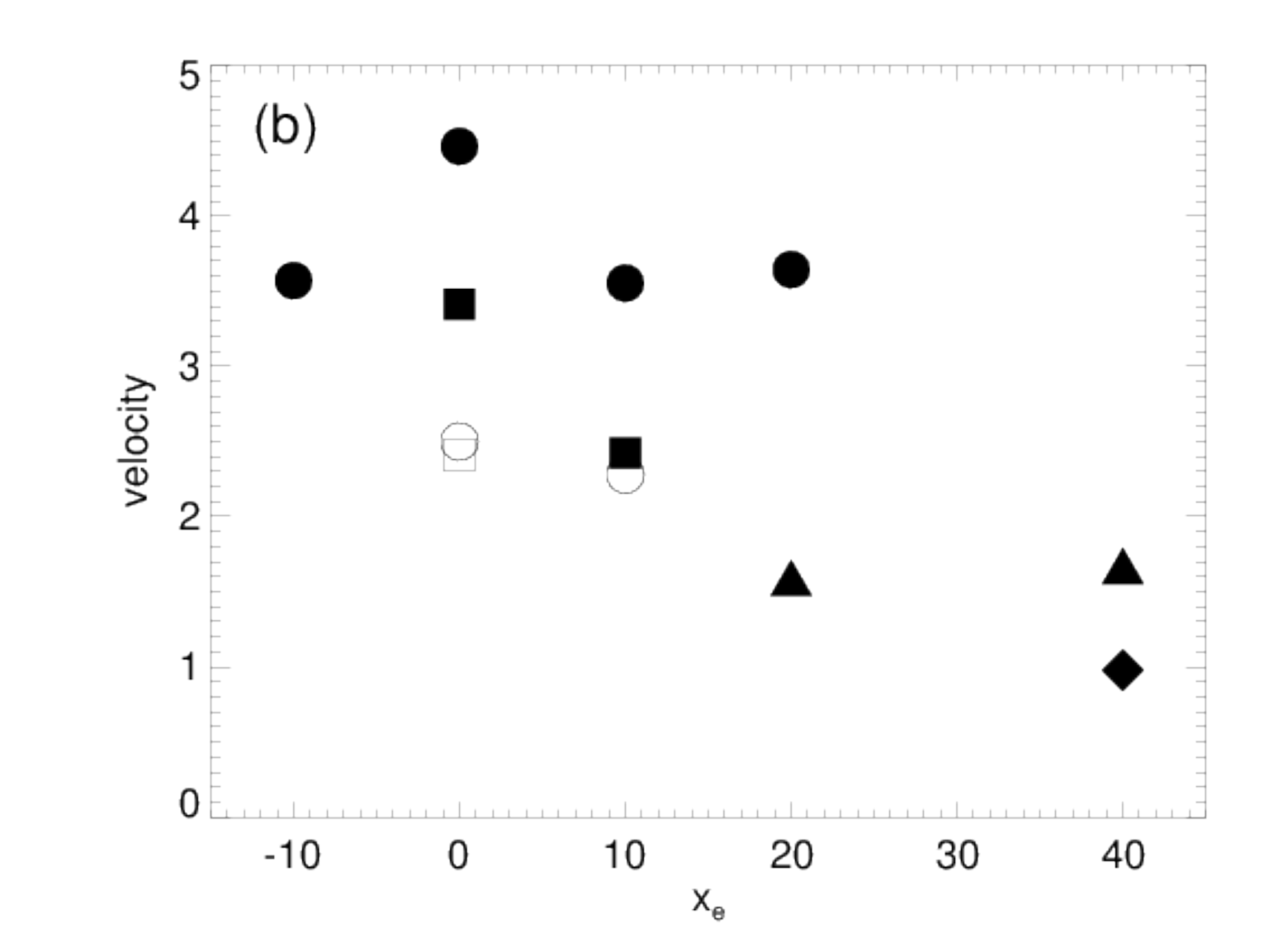}
      \end{minipage}
    \end{tabular}
  \end{center}
  \caption{\small{(a): Time evolution of height of the plasmoid in each eruptive 
      case. Solid lines represent $B_{e}=17$ cases, dashed lines represent 
      $B_{e}=15$ cases, dot-dashed lines represent $B_{e}=13$ cases, and the  dotted line 
      represents $B_{e}=11$ case. The parentheses indicate $(B_{e},x_{e})$.
      We show only the cases with a=100.
      (b): mean ejection speed of the plasmoid in each eruptive case. 
      Circles($\circ $) represent $B_{e}=17$ cases, squares($\Box $) 
      represent $B_{e}=15$ cases, triangles($\triangle $) represent $B_{e}=13$ cases, 
      and diamonds($\Diamond $) represent $B_{e}=11$ case. The filled symbols represent
      the erupted cases, and the open symbols represent the formed cases (confined eruption).
  }}
  \label{height}
\end{figure}

Fig.\ref{height} (a) shows the time evolution of height, 
and Fig.\ref{height} (b) shows the mean ejection speed of the 
plasmoid in each eruptive and confined case.
The temporal ascent and subsequent descent of the plasmoid 
from the co-arranged arcade in the cases 
of $(B_{e},x_{e},a)=(15,0.100), (17,0,100)$ and that
from the counter-arranged arcade in the case of $(B_{e},x_{e},a)=(17,10,100)$
are regarded as the confined eruptions.
The mean ejection speed of each plasmoid is calculated from the inclination 
of the time-height curves. 
Because the confined eruptions have not only the ascending phase but also
the subsequent descending phase of plasmoid, 
we use the time-height curve only during the ascending phase for the
computation of the mean ejection speed.

When the field strength of the emerging flux is large, the mean ejection speed is high.
The dependence on $x_{e} $ is relatively small.
This means that the kinetic energy of the plasmoid comes from the magnetic energy
of the emerging flux, i.e., it is stored
in the corona through the deformation of the coronal arcade field by the emerging flux
and the reconnection releases the
stored magnetic energy as kinetic energy of the plasmoid.

The mean ejection speed of the confined eruptions 
(the open symbols in Fig.\ref{height} (b)) 
is smaller than those of full eruptions in the same situations.
This is due to the downward magnetic tension force of overlying arcade field lines, 
which is formed by the reconnection between the co-arranged arcade 
and the emerging flux (see also Fig.\ref{ca_mecha}).

\section{DISCUSSION}
In this section, we discuss the reasons why the results of the parameter survey show 
the tendencies in Figs.\ref{resr} and \ref{resa} and attempt to outline the  eruptive conditions of each mechanism.
\subsection{Conditions for CA-Type Eruptions}
When the flux tube emerges around the edges of the arcade fields, 
the expanding emerging flux compresses the arcade fields, 
which generates a vertical current sheet above the PILs. 
If the current exceeds a certain threshold, reconnection occurs,
resulting in the formation and eruption of a plasmoid. 
Therefore, deriving the eruptive conditions requires the evaluation of 
the current density inside the compressed arcade field, 
of the extent to which the emerging flux 
compresses the arcade field, and of the threshold of the current 
to trigger the reconnection. 
Though our simulation set-up include the multiple arcades 
compressed by the emerging flux, 
we discuss the eruptive condition by using the system which has 
one arcade and one emerging flux tube 
in order to simplify the problem and to get fundamental physical picture.

\begin{figure}
  \begin{center}
    \includegraphics[bb=0 0 516 210,scale=0.8]{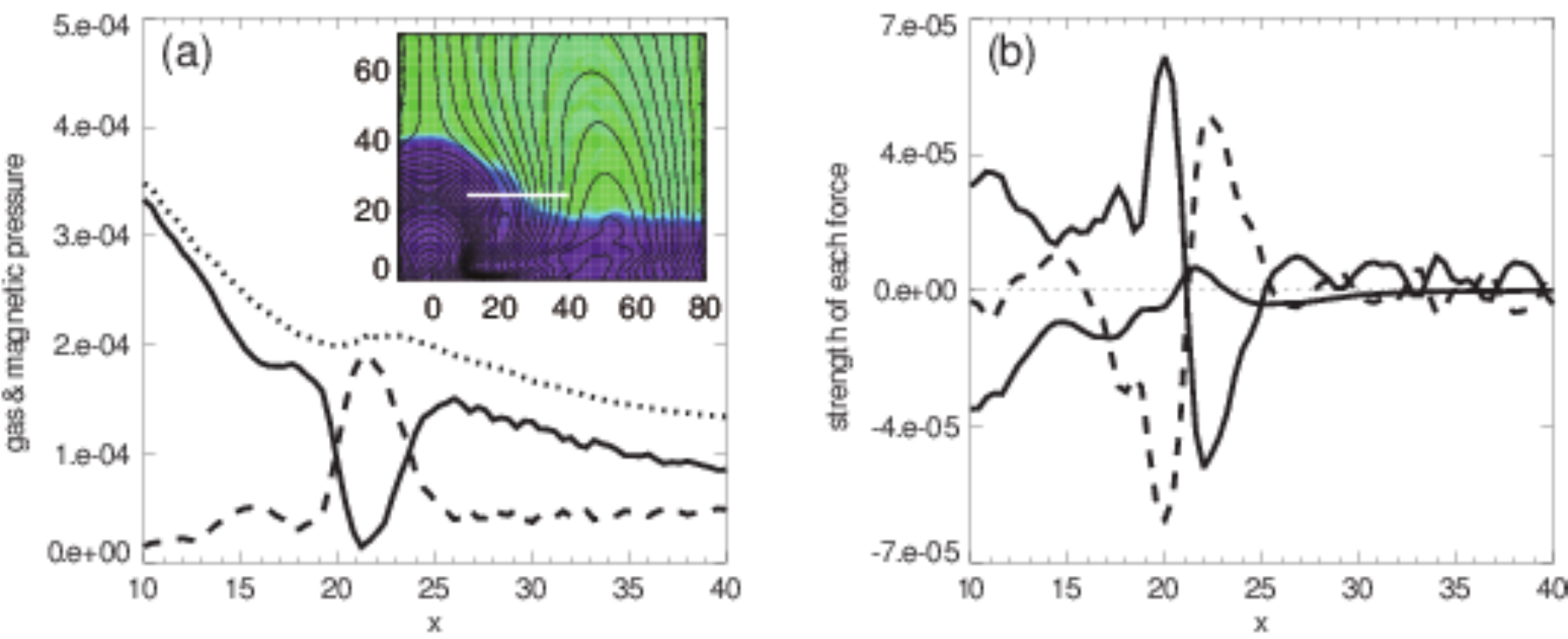}
  \end{center}
  \caption{(a): Pressure in the vicinity of  emerging flux.
    Solid line is magnetic pressure, dashed line is gas pressure 
    and dotted line is total pressure along the white horizontal line in the inset. 
    The inset is the enlarged view of t=90.0 in Fig.\ref{unformed}.
    (b): $x$-directional forces at the same region as (a).
    Solid line is magnetic pressure gradient, dashed line is gas pressure gradient
    and dot-dashed line is magnetic tension force.}
  \label{each_force}
\end{figure}

First, we evaluate the extent to which the emerging flux compresses 
the arcade field. This can be estimated by considering the balance of 
magnetic pressures between the arcade and emerging flux.
Fig.\ref{each_force} (a) shows magnetic pressure (solid line), 
gas pressure (dashed line), and  total pressure (dotted line) on the
white line in the inset corresponding to Fig.\ref{unformed} 
(the case of $(B_{e},x_{e},a)=(11,0,100)$, $t=90.0$).
Fig.\ref{each_force} (b) shows the $x$-directional forces of the magnetic pressure gradient, 
gas pressure gradient, and magnetic tension force.
The region between $x \approx 18$ and $x \approx 25$, next to the interface between the 
emerging flux and the arcade, is a high-$\beta $ region.
The high-$\beta $ region is composed of the material fallen from the top of the emerging flux
where `dense sheath' (\citealp{2006PASJ...58..423I}) is formed. 
As shown in Fig.\ref{each_force} (a) and (b), 
the magnetic pressure gradients of the emerging flux and the arcade field are balanced 
with the gas pressure gradient of the high-$\beta $ region. 
According to Fig.\ref{each_force} (b),
the magnetic tension force is negligible compared with the other two forces.
Therefore, it is possible
to regard the magnetic pressures of the emerging flux and the arcade
field as being in balance with each other.
Next, we postulate the self-similar time evolution of the magnetic fields, 
assuming that the parameters of the arcade field $(B_{a},L,a)$ are changed 
into $(\bar{B_{a}},fL,\bar{a})$ after being compressed. 
The extent that the emerging flux can compress the arcade is expressed by 
the compression rate of the arcade field $f$.
From the conservation of  magnetic flux,
\begin{eqnarray*}
\Phi _{a} = \int _{0} ^{L} B_{a}\cos \left( \frac{\pi }{2L}x \right)
\exp \left[ -\frac{y}{a} \right]dx 
&=&\int _{0} ^{fL} \bar{B_{a}} \cos \left( \frac{\pi }{2fL}x\right)
\exp \left[ -\frac{y}{\bar{a}} \right]dx,\\
\end{eqnarray*}
which leads to
\begin{equation}
\Phi _{a} = \frac{2L}{\pi }B_{a}\exp [-\frac{y}{a}] 
= \frac{2fL}{\pi }\bar{B_{a}}\exp [-\frac{y}{\bar{a}}] . \label{fluxa}
\end{equation}
The condition 
\begin{equation}
\bar{B_{a}}=\frac{B_{a}}{f},~~\bar{a}=a \label{babar}
\label{afterc}
\end{equation}
satisfies Eq.(\ref{fluxa}). Actually, $\bar{a}=a$ is just an assumption, 
but when $y$ is small enough compared with $a$, the effect from a change in $a$ is negligible,
which is confirmed in our results.
For the  emerging flux, $(B_{e},q,r_{e})$ changed
into $(\bar{B_{e}},\bar{q},\bar{r_{e}})$ after compression. The expanded emerging flux is described as
\begin{equation}
  \bar{\vctr{B_{e}}}=\bar{B_{\theta }}\vctr{e}_{\theta }+\bar{B_{z}}\vctr{e}_{z},
\end{equation}
where
\begin{eqnarray}
  \bar{B_{\theta }}&=&\bar{q}r\bar{B_{z}}, \\
  \bar{B_{z}}&=&\bar{B_{e}}F(r),
\end{eqnarray}
and where $\vctr{e}_{\theta }$ and $\vctr{e}_{z}$ are unit vectors of azimuthal direction
and $z$-direction, respectively, and
$F(r)$ is an arbitrary decreasing function of $r$ which satisfies $F(0)=1$.
From the conservation of magnetic flux,
\begin{equation}
  \Phi _{e}=\int _{0}^{r_{e}}\frac{B_{e}qr}{1+q^{2}r^{2}}dr
  =\int _{0}^{\bar{r_{e}}}\bar{B_{e}}\bar{q}rF(r)dr, 
\end{equation}
which leads to
\begin{equation}
  \Phi _{e}=\frac{B_{e}}{2q}\ln (1+q^{2}r_{e}^{2}) 
  = \bar{B_{e}}\bar{q}r_{e}^{2}G(\bar{r_{e}}), \label{fluxe}
\end{equation}
where
\begin{equation}
  G(\bar{r_{e}})\equiv \frac{1}{\bar{r_{e}}^{2}}\int _{0}^{\bar{r_{e}}} rF(r)dr.
\end{equation}
Because $F(r)$ is a monotonically decreasing function of $r$, 
\begin{equation}
\frac{1}{2}F(\bar{r_{e}}) \leq G(\bar{r_{e}})  .
\label{fgneq} 
\end{equation}
The emerging flux can expand until its magnetic pressure is balanced with 
that of the arcade field:
\begin{equation}
\frac{(1+\bar{q}^{2}\bar{r_{e}}^{2})\bar{B_{e}}^{2}F(r)^{2}}{8\pi }
=\frac{\bar{B_{a}}^{2}}{8\pi }\exp \left[ -\frac{2y}{\bar{a}}\right] .
\label{balance}
\end{equation}
By substituting Eq.(\ref{fluxa}) and Eq.(\ref{fluxe}) into Eq.(\ref{balance}) and using
the approximation that $\bar{q}\bar{r_{e}} \gg 1$,  
\begin{equation}
  \frac{\Phi _{e}}{\Phi _{a}}
  =\frac{\pi \bar{r_{e}}}{2fL}\frac{G(\bar{r_{e}})}{F(\bar{r_{e}})} .
  \label{sq1}
\end{equation}
From the positional relation between the emerging flux and the arcade field,
for the counter-arranged arcade field, as $\chi L \equiv L-x_{e}$,
\begin{equation}
  2L = x_{e}+\bar{r_{e}}+f(2L-x_{e}), 
\end{equation}
and for the co-arranged arcade field, as $\chi L \equiv x_{e}+L$,
\begin{equation}
  -2L = x_{e}-\bar{r_{e}}-f(x_{e}+2L),
\end{equation}
which leads to
\begin{equation} 
  \bar{r_{e}} = (1-f)(1+\chi )L , \label{distance1}  
\end{equation}
where $\chi L$ represents the absolute distance between the PIL of the arcade field 
and the location of the emerging flux.
By using Eqs.(\ref{fgneq}), (\ref{sq1}), and (\ref{distance1}),
\begin{equation}
  \frac{\zeta }{1+\chi } > \frac{\pi }{4} \frac{1-f}{f} ,
\end{equation}
where $\zeta $ is the ratio of the emerging flux to the arcade flux defined as
\begin{equation}
  \zeta = \frac{\Phi _{e}}{\Phi_{a}},
  \label{zeta0}
\end{equation}
and $\chi $ is the distance between the PIL of the arcade field and the location of the
flux emergence normalize by the half of the 
width of the arcade,
\begin{equation}
  \chi = \frac{L-x_{e}}{L}, \label{loc_counter}
\end{equation}
for the counter-arranged cases and
\begin{equation}
  \chi = \frac{x_{e}+L}{L}. \label{loc_co}
\end{equation}
for the co-arranged cases.
From these formulas, the extent of compression
(in this case minimum value of $f$) can be estimated.
By assuming that reconnection occurs when $f<f_{\tani{cr}}$, the necessary condition
for triggering an eruption is
\begin{equation}
  \frac{\zeta }{1+\chi } > \frac{\pi }{4} \frac{1-f_{\tani{cr}}}{f_{\tani{cr}}} .
\end{equation}
If the force-free condition, 
$(\nabla \times \bar{\vctr{B_{e}}}) \times \vctr{\bar{B_{e}}}=0 $,
is satisfied, the function $F(r)$ is derived 
to be the Gold-Hoyle type geometry: 
\begin{equation}
  \frac{d}{dr}\left\{ \left(1+q^{2}r^{2} \right) F(r)\right\} = 0, 
\end{equation}
from which we obtain
\begin{equation} 
  F(r)=\frac{1}{1+q^{2}r^{2}}.
\end{equation}
When a static state is realized due to the balance of magnetic pressures, 
the emerging flux can be assumed to be under a force-free condition.
In this case, the eruptive conditions are easily derived by using Eq.(\ref{sq1}), 
Eq.(\ref{distance1}) and 
 the approximation of $\bar{q}\bar{r_{e}} \gg 1$:\begin{equation}
  \frac{\zeta }{1+\chi } > \kappa \frac{1-f_{\tani{cr}}}{f_{\tani{cr}}}, 
  \label{econd0}
\end{equation}
where
\begin{equation}
  \kappa = \frac{\pi }{4} \ln(\bar{q}^{2}\bar{r_{e}}^{2}) .
\end{equation}
Because of the logarithmic dependence, the value of $\kappa $ can be regarded as constant 
even when the value of $\bar{r_{e}}$ varies. 
The value of $\kappa $ is $3.6$ to $5.3$ when $\bar{r_{e}}$ is $20$ to $60$ 
in our simulations.

Next, we discuss the critical value for compression rate $f_{\tani{cr}}$. 
We use anomalous resistivity, which depends on the drift velocity
$v_{d}=mJ/(e\rho )$ (see Eq.(\ref{resis})).
The current density inside the arcade field in the initial state $J$ is 
\begin{equation}
  J=\sqrt{\left( \frac{\pi }{2L}\right)^{2}-\left( \frac{1}{a}\right)^{2}}
  B_{a}\exp \left[ -\frac{y}{a}\right]
  =\frac{\pi }{2L}\sin \theta B_{a}\exp \left[ -\frac{y}{a}\right] .
\end{equation}
The current density inside the compressed arcade $\bar{J}$ is
\begin{equation}
  \bar{J}
  =\frac{\pi }{2fL}\sin \bar{\theta }\bar{B_{a}}\exp \left[ -\frac{y}{\bar{a}}\right] 
  =\frac{J}{f^{2}},
\end{equation}
where Eq.(\ref{babar}) and the approximation of
\begin{equation}
  \frac{\sin \bar{\theta }}{\sin \theta }
  =\sqrt{\frac{(\pi a)^{2}-(2fL)^{2}}{(\pi a)^{2}-(2L)^{2}}}\approx 1
\end{equation}
are used. The drift velocity $\bar{v_{d}}$ inside the compressed arcade is
\begin{equation}
\bar{v_{d}}=\frac{m\bar{J}}{e\bar{\rho }}=\frac{v_{d}}{f},
\end{equation}
where the density inside the compressed arcade is assumed to be $\bar{\rho }=\rho /f$.
Assuming that reconnection occurs when the drift velocity exceeds 
a certain critical value, namely $\bar{v_{d}} > v_{\tani{cr}}$, then 
\begin{equation}
  f<\frac{v_{d}}{v_{\tani{cr}}}=f_{\tani{cr}}\label{fcr}
\end{equation}
is the condition for $f$ to trigger the reconnection. By combining
Eq.(\ref{econd0}) and Eq.(\ref{fcr}), the eruptive condition is
\begin{equation}
  \frac{\zeta }{1+\chi } > \kappa ^{\prime },
  \label{econd}
\end{equation} 
where
\begin{equation}
  \kappa ^{\prime }=\kappa \frac{v_{\tani{cr}}-v_{d}}{v_{d}} .
  \label{kappap}
\end{equation}
The theory of resistivity in the corona is still unclear, so 
the dependence on $v_{d}$ and other parameters might be more complex. 
This model of $\kappa ^{\prime }$ is used only for a qualitative explanation for 
our simulation results. 

\begin{figure}
  \begin{center}
    \includegraphics[bb=0 0 425 283,scale=0.7]{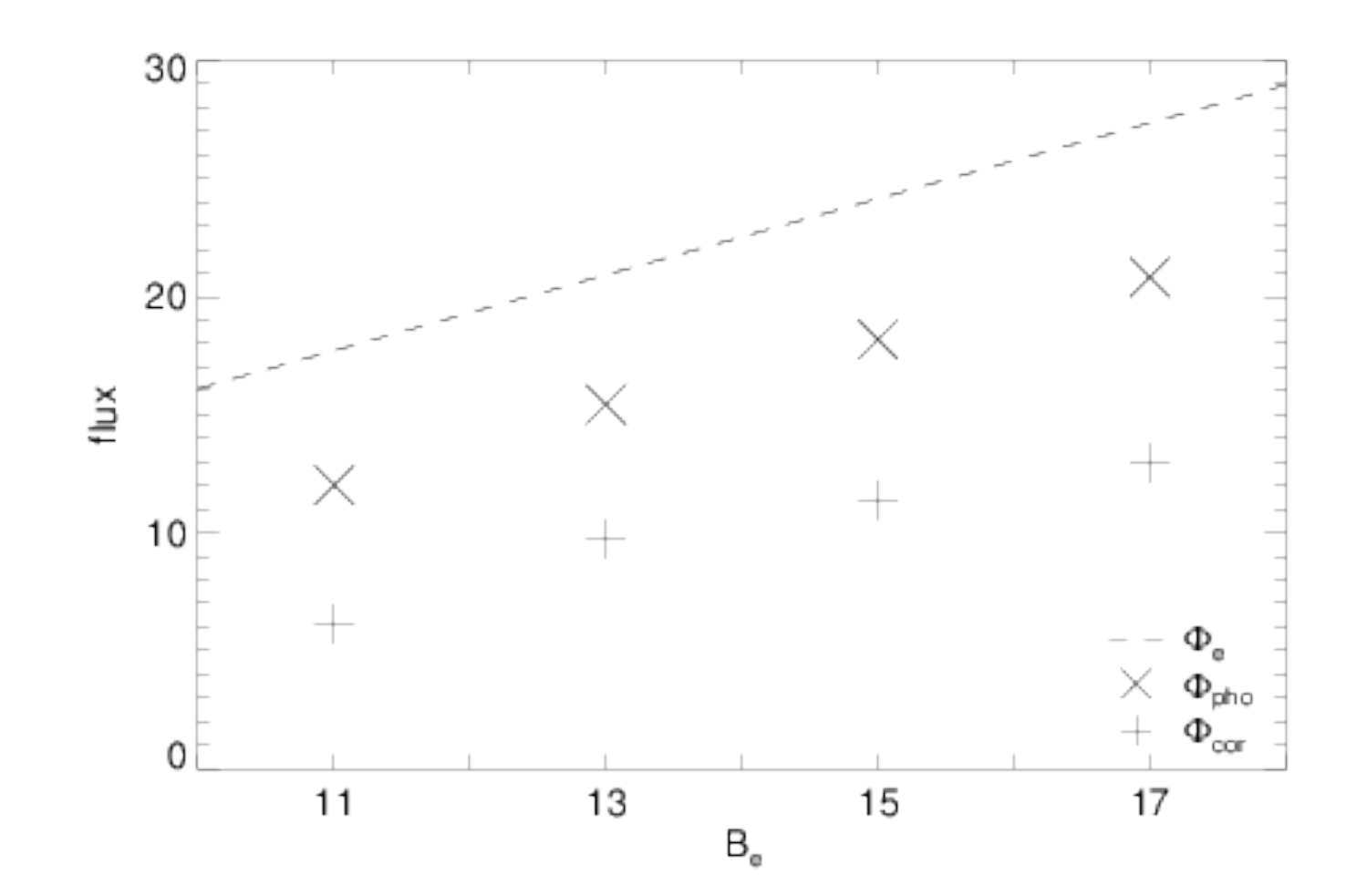}
    \caption{\small{
        Amount of magnetic flux injected into photosphere and corona.
        $\times $ is $\Phi _{\tani{pho}}$  and $+$ is $\Phi _{\tani{cor}}$. Dashed 
        line represents the initial magnetic flux of the sub-photospheric flux tube.} }
    \label{fluxpt}
  \end{center}  
\end{figure}

\begin{figure}
\begin{tabular}{c}
  \begin{minipage}{1.0\hsize}
    \begin{center}
      \includegraphics[bb=0 0 850 566,scale=0.4]{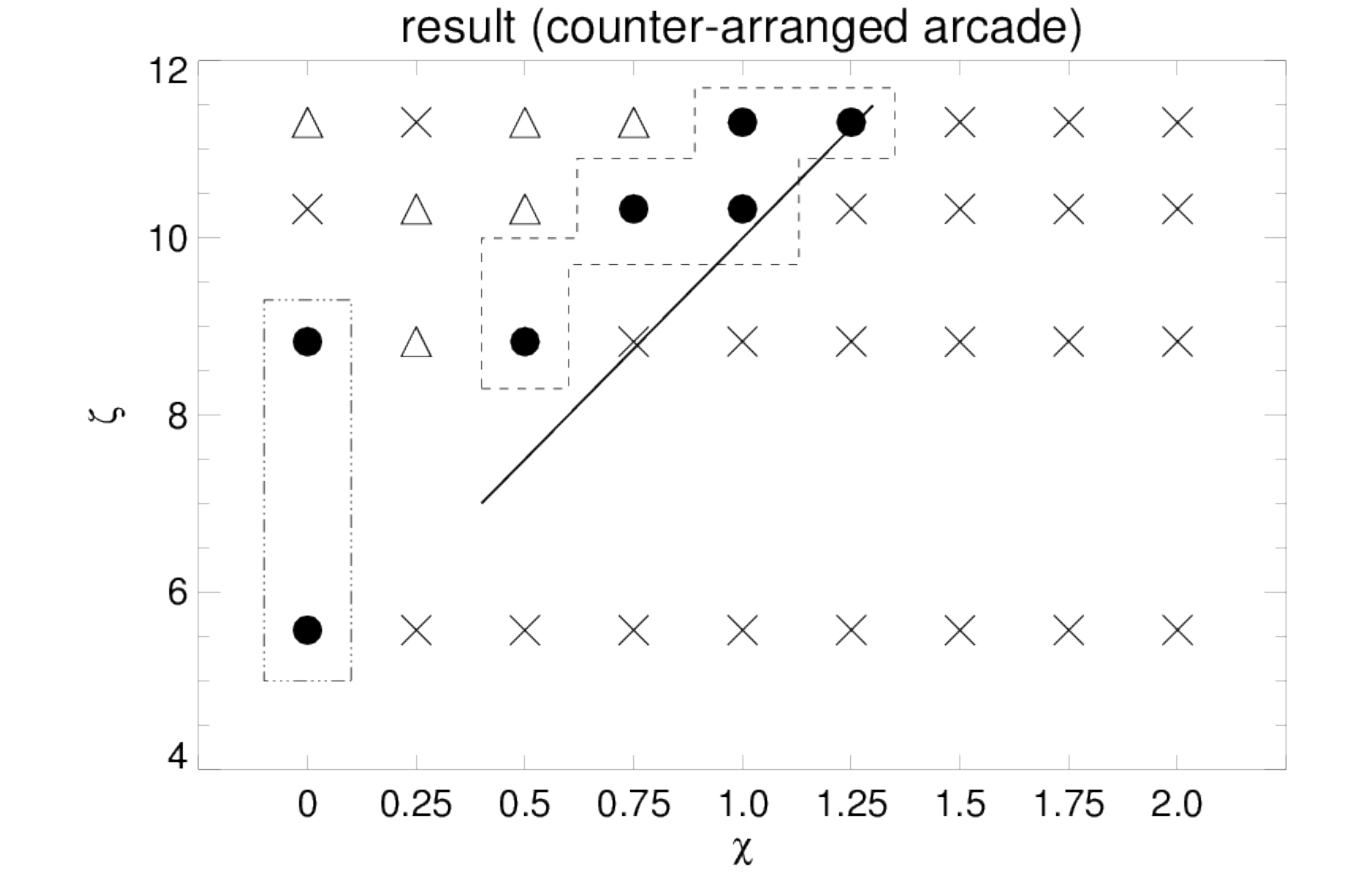}
    \end{center}
  \end{minipage}
  \\
  \\
  \begin{minipage}{1.0\hsize}
    \begin{center}
      \includegraphics[bb=0 0 850 566,scale=0.4]{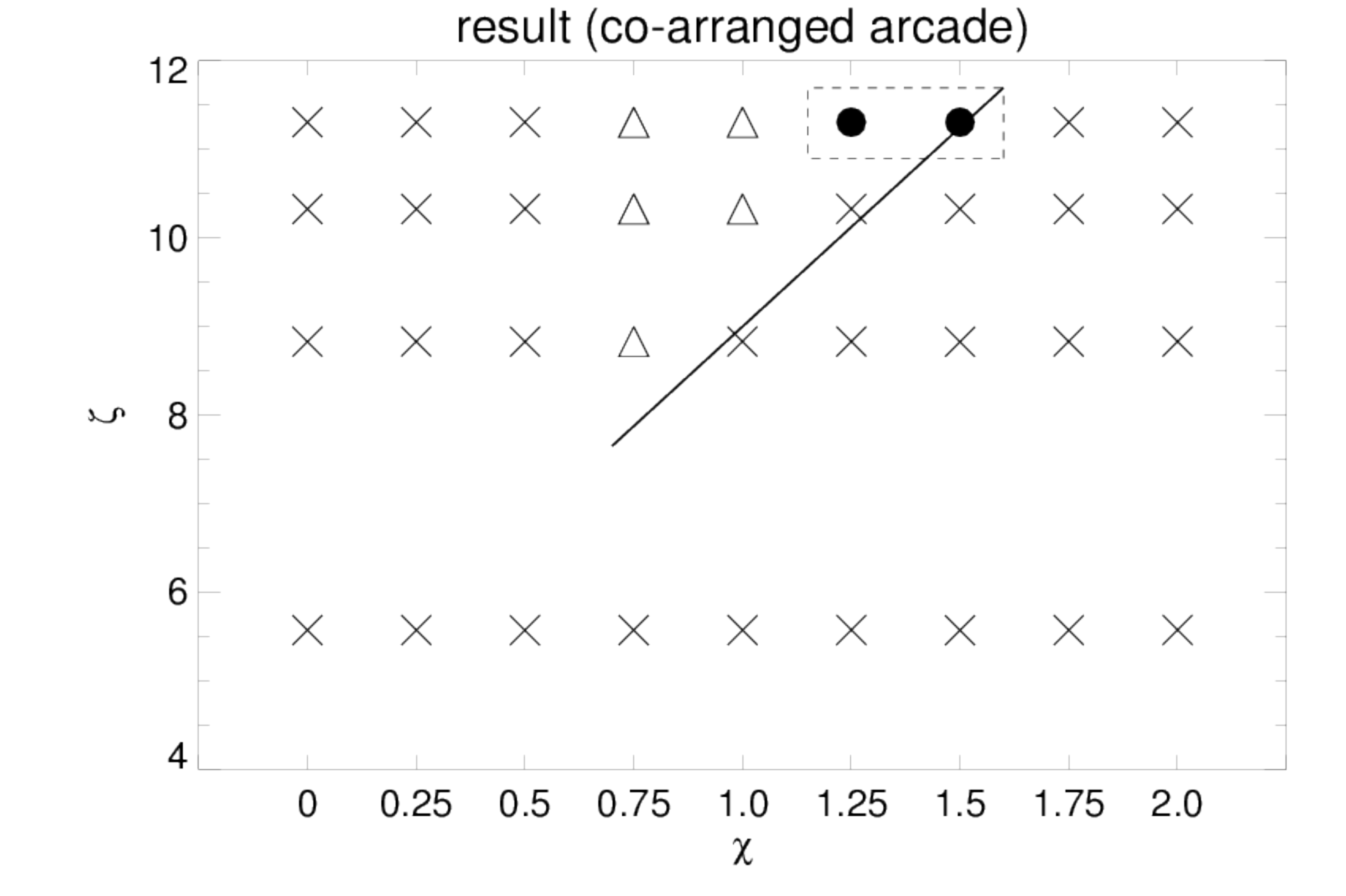}
    \end{center}
  \end{minipage}
  \end{tabular}
  \caption{\small{
      Rearrangement of the results of parameter survey shown in Fig.\ref{resr}. 
      The upper panel corresponds to the results of the counter-arranged arcades, 
      and the lower panel corresponds to that of the co-arranged arcades.
      Circles are `erupted' cases, triangles are `formed' cases, and crosses 
      are `unformed' cases. $\zeta $ is the ratio of the emerging flux 
      and the arcade flux as defined in Eq.(\ref{zetae}). 
      $\chi $ is the normalized distance between the
      PIL of the either arcade and the location of the emerging flux defined 
      as Eq.(\ref{loc_counter}) and (\ref{loc_co}).
      $\chi \leq 1.0$ is inner arcade region and $\chi \geq 1.0$ is outer arcade region.
      Circles in dashed boxes represent  CA-type eruptions and those in dot-dashed boxes 
      represent  RC-type eruptions. Solid lines represent the eruptive condition expressed
      by formula (\ref{econd}).
  }}
  \label{cocounter}  
\end{figure}

The obtained model can be compared with our simulation results.
For this purpose, $\zeta $, $\Phi _{e}$, and $\Phi _{a}$ are measured 
from the simulations. Fig.\ref{fluxpt} shows the amount of 
the magnetic flux introduced  into the photosphere and the corona.
The dashed line in Fig.\ref{fluxpt} is the initial magnetic flux of the emerging flux 
computed by the middle term of Eq.(\ref{fluxe}).
The amount of magnetic flux introduced into the photosphere and the corona
is defined as follows:
\begin{eqnarray}
  \Phi _{\tani{pho}}(t) &=& \frac{1}{2} \left(\Phi (y_{\tani{pho}},t) - \Phi _{\tani{ar}} 
  (y_{\tani{pho}})\right), \\
  \Phi _{\tani{cor}}(t) &=& \frac{1}{2} \left(\Phi (y_{\tani{cor}},t) - \Phi _{\tani{ar}} 
  (y_{\tani{cor}})\right), \\
\end{eqnarray}
where
\begin{eqnarray}
  \Phi (y,t)&=&\int _{x_{\tani{min}}} ^{x_{\tani{max}}} |B_{y}(x,y,t)|dx, \\
  \Phi _{\tani{ar}}(y)&=&\int _{x_{\tani{min}}} ^{x_{\tani{max}}} |B_{y}(x,y,t=0)|dx.
\end{eqnarray}
The total amount of the initial magnetic flux of the arcade field 
in the whole region is $\Phi _{\tani{ar}}$, so that $\Phi _{\tani{ar}}=8\Phi _{a}$ 
(note that there are four arcades in the simulation box and $\Phi _{a}$ means the 
amount of flux of one polarity). The values of
$\Phi _{\tani{pho}} $ and $\Phi _{\tani{cor}}$ 
in Fig.\ref{fluxpt} are maximum values. 
We compute the value of $\zeta $ by using $\Phi _{\tani{cor}}$ instead of $\Phi _{e}$ and 
$\Phi _{\tani{ar}}(y_{\tani{cor}})/8$ instead of $\Phi _{a}$ in Eq.(\ref{zeta0}).
\begin{equation}
  \zeta =\frac{\Phi _{e}}{\Phi _{a}}=\frac{\Phi _{\tani{cor}}}{\Phi _{\tani{ar}}/8}. 
  \label{zetae}
\end{equation}
The condition expressed by formula (\ref{econd}) can well explain 
the transition between the eruptive cases (including the formed cases) 
and the unformed cases shown in Fig.\ref{cocounter}
by adopting a proper value of $\kappa ^{\prime }$.
The solid line in Fig.\ref{cocounter} represents the criterion for  formula (\ref{econd}) 
of $\kappa ^{\prime }=5.0$ 
for the counter-arranged case and $\kappa ^{\prime }=4.5$ for the co-arranged case. As shown in Fig.\ref{cocounter}, the condition expressed by formula (\ref{econd}) can well explain the results of parameter surveys for CA-type eruptions originating from both the co-arranged and counter-arranged arcades. 

The reduction of $\kappa ^{\prime }$ for the co-arranged case compared with 
the counter-arranged case can be interpreted as follows.
Because of the diffusion (reconnection) with the emerging flux, 
the co-arranged arcade loses its magnetic energy
during compression, which leads to a reduction in its magnetic pressure.
This means the co-arranged arcade is more easily compressed, 
and reconnection inside the arcade is more easily triggered, 
compared with the counter-arranged arcade.

Fig.\ref{resa} shows that, when the magnetic scale height $a$ is large,
eruptions are more likely to occur.
When $a$ is large, the current density
is high, causing  the initial drift velocity $v_{d}$ to be large. 
The dependence of $\kappa ^{\prime }$ on
$v_{d}$ in Eq.(\ref{kappap}) shows that eruptions are more likely 
to occur when $v_{d}$ is large because of the low value of $\kappa ^{\prime}$. 
Therefore, the obtained model can also qualitatively explain the simulation results regarding the
magnetic scale height $a$. 

In a three-dimensional case, it is probable that
 larger values of $\kappa $ are required because the radius of expanded emerging 
flux is smaller than that in a two-dimensional case.
In fact, the magnetic field strength of the emerging flux that caused an eruption in \cite{Notoya2007ASPC} was twice as large as that in our results.

\subsection{Conditions for RC-Type Eruptions}
When the  emerging flux is located at the PIL, 
it does not lead to a sideways compression of the arcade and the formation of a vertical 
current sheet. 
Hence, CA-type eruptions are not likely to occur.
In the case of  counter-arranged arcades, some plasmoids are formed through
reconnections in the long current sheet between 
the arcade and the emerging flux. 
The reconnection coupling, the linkage of outflow of one reconnection to inflow
of another reconnection, enables a plasmoid to erupt.
The formation of  small plasmoids is always seen in the 
current sheet between the arcade and the emerging flux.
On the other hand, it is not always predictable from our simulations
whether the plasmoids will grow to  large plasmoids or not. 
In addition,
when  reconnection coupling does not occur, an eruption does not occur, 
even if there is a continuous growth of a plasmoid.   
The tendency seen in the results of the parameter survey (Fig.\ref{resr}) indicates 
that RC-type eruptions
occur only when the field strength of the emerging flux $B_{e}$ is small.
This is because the reconnection coupling is likely to occur when the two reconnection 
points are close enough to link with each other. The distance between the two 
reconnection points depends on the expansion radius of the emerging flux, 
so that a smaller field strength of  emerging flux is preferable.
When the location of the emerging flux surrounds the PIL of
the co-arranged arcade,  no current sheet forms between the arcade and 
emerging flux, resulting in an `unformed' case.  

RC-type eruptions  and `reversed shear' (RS)-type eruptions, as observed in \cite{2012ApJ...760...31K}, 
occur under similar conditions.
Our simulations reveal that the eruptions are likely to occur 
when the magnetic scale height $a$ is large (Fig.\ref{resa}).
\cite{2012ApJ...760...31K} revealed that their RS-type eruptions are triggered when
the shear angle of the arcade is large.
According to the relation between $a$ and $\theta $ in Eq.(\ref{dtheta}), 
for constant $L$, larger values of $a$ correspond to larger 
values of $\theta $, meaning that our  parameter dependence on $a$  qualitatively
agrees with the results in \cite{2012ApJ...760...31K}.
One difference is the formation process of the plasmoids.
In RS-type eruptions in \citet{2004ApJ...610..537K,2012ApJ...760...31K}, 
the force that causes  the magnetic field lines of arcades to converge, by which reconnection is 
triggered, is  the magnetic pressure gradient. In our simulations, the magnetic 
pressure gradient is immediately balanced with the gas pressure gradient.  
Instead of the pressure gradient, the convergence of field lines is enhanced by the coupling of multi-point 
reconnections in our RC-type eruptions.

\subsection{Formed Case}
Formed cases, in which the plasmoid is formed but non-eruptive, 
are shown with triangles in 
Fig.\ref{resr}, \ref{resa} and \ref{cocounter}.
In these cases, the eruption of a plasmoid that forms in a
co-arranged arcade is suppressed by the stabilizing flux of
the overlying arcade, which successively increases due to 
reconnection between the arcade and the emerging flux 
(see bottom pannels in Fig.\ref{ca_mecha}).
Another reason for non-eruption is that the earlier eruption from either 
the counter- or co-arranged arcade 
blocks the passage of the later eruption. This can happen in the cases of CA-type eruptions 
if the emerging flux is strong enough to trigger reconnection in both arcades.
No eruptions are seen also when the coupling of two reconnection points fails. 
This is likely for the plasmoids formed in RC-type condition with strong emerging flux.  
Even when the plasmoids grow sufficiently, 
they do not erupt but remain stationary without the reconnection coupling.
The continuous growth of plasmoids in formed cases might lead to eruptions 
in three-dimensional simulations by the help of the kink or torus instability.

\subsection{Comparison with Other Studies}
\cite{1995JGR...100.3355F} first showed in their statistical study 
that eruptions were triggered when the spatial arrangement of pre-existing 
arcade fields and newly emerging flux was favorable for reconnection. 
In our results, the RC-type eruptions are clearly consistent 
with their observation.
In the case of CA-types, both favorable and unfavorable reconnection arrangements 
can cause eruptions. This is because the most essential point 
for  CA-type eruptions 
is whether or not the arcade field is compressed enough to cause reconnection. 
\citet{2008ChA&A..32...56X} statistically investigated the triggering mechanism 
of eruptions caused by emerging flux, especially with respect to the 
polarity, amount and the location of emerging flux. 
They reported that eruptions did not occur when the emerging flux had same polarity orientation
as that of the ambient magnetic field and was located too close to or too far from the PIL, 
and when the emerging flux is too weak. Our results are consistent with their findings.
They also showed that eruptions were triggered when the emerging flux 
had opposite polarity orientation and was located inside the filament channel, or
when the emerging flux had same polarity orientation and was located outside the filament channel.
The former case agrees with our RC-type conditions.
The latter case corresponds to our CA-type condition with the co-arranged arcade and
does not conflict with our results, while the CA-type eruptions are also triggered 
from the counter-arranged arcades in our simulations.

\cite{Lin2001JGR} theoretically investigate the condition for the eruptions by interaction
between the newly emerging flux and the pre-existing coronal magnetic field which
contains a flux rope. Their analysis is based on the loss of equilibrium model. 
Our results of parameter survey agree with one of their results that
the eruptions can occur even when the orientation of the emerging flux
and the coronal magnetic field is unfavorable for reconnection. 
One difference between theirs and ours lies in the initial coronal magnetic field. 
In our case, the flux rope (plasmoid) is formed
after the interaction between the emerging flux and the coronal arcade field.
In the case of \cite{Lin2001JGR}, the initial coronal magnetic field contains the flux rope.
The other difference is that our simulations solve the flux emergence
self-consistently. In the study of \cite{Lin2001JGR}, the parameters of
emerging flux such as the field strength and the vertical location are independent with
each other. In addition, the flux rope contained in the initial coronal field also 
has several independent parameters. Because of these many independent parameters, 
the conditions for eruptions became complicated.
In the present study, because the flux emergence is solved self-consistently
and the initial coronal field has simple arcade configuration,
the number of the independent parameters are reduced.
This makes the physical nature and the eruptive condition more simple.

Our simulations are applicable to the eruptions triggered by flux emergence in
so-called pseudostreamers, since two neighboring arcades in our system correspond 
topologically to the two closed flux lobes at the base of a pseudostreamers.
\citet{Lynch2013ApJ} modeled the sympathetic eruptions
by imposing the shearing flow at the two PILs of the pseudostreamer.
In their simulations, both the two arcades of pseudostreamer are erupted.
It is common between our CA-type mechanism  
and their mechanism
that the reconnections are triggered inside each arcade,
leading to the formation and eruption of the plasmoid.
The difference is that our model does not result in the sympathetic eruptions.
One reason is that the eruptions from the co-arranged arcade 
tend to be confined by the large overlying arcade, which is formed 
through the reconnection between the emerging flux and the co-arranged arcade.
The other reason is that the earlier eruption from either the counter-arranged or the 
co-arranged arcade blocks the path of the later eruption.

\begin{figure}
\begin{tabular}{c}
  \begin{minipage}{\hsize}
    \begin{center}
      \includegraphics[bb=0 0 708 340,scale=0.45]{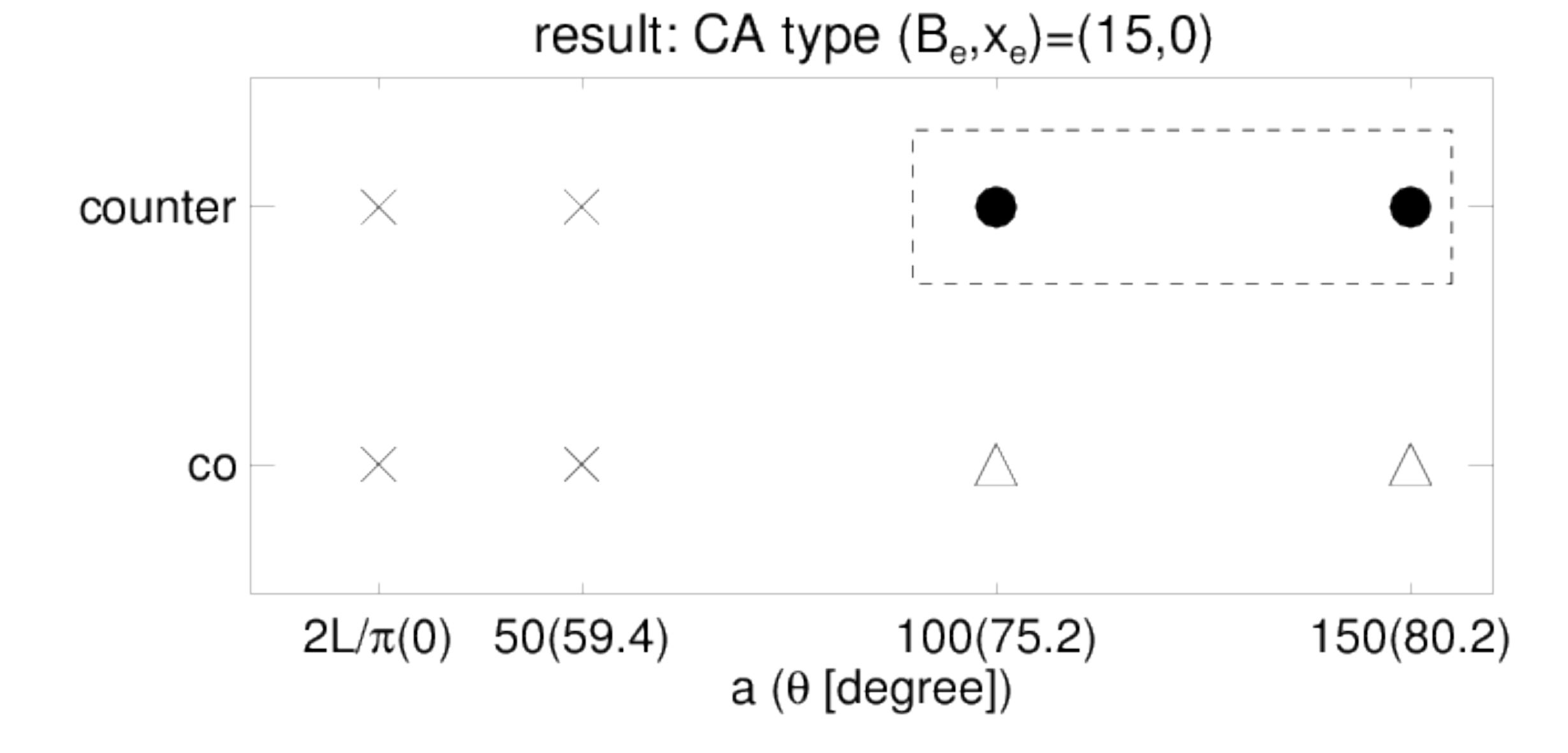}
    \end{center}
  \end{minipage}
  \\
  \begin{minipage}{\hsize}
    \begin{center}
      \includegraphics[bb=0 0 708 340,scale=0.45]{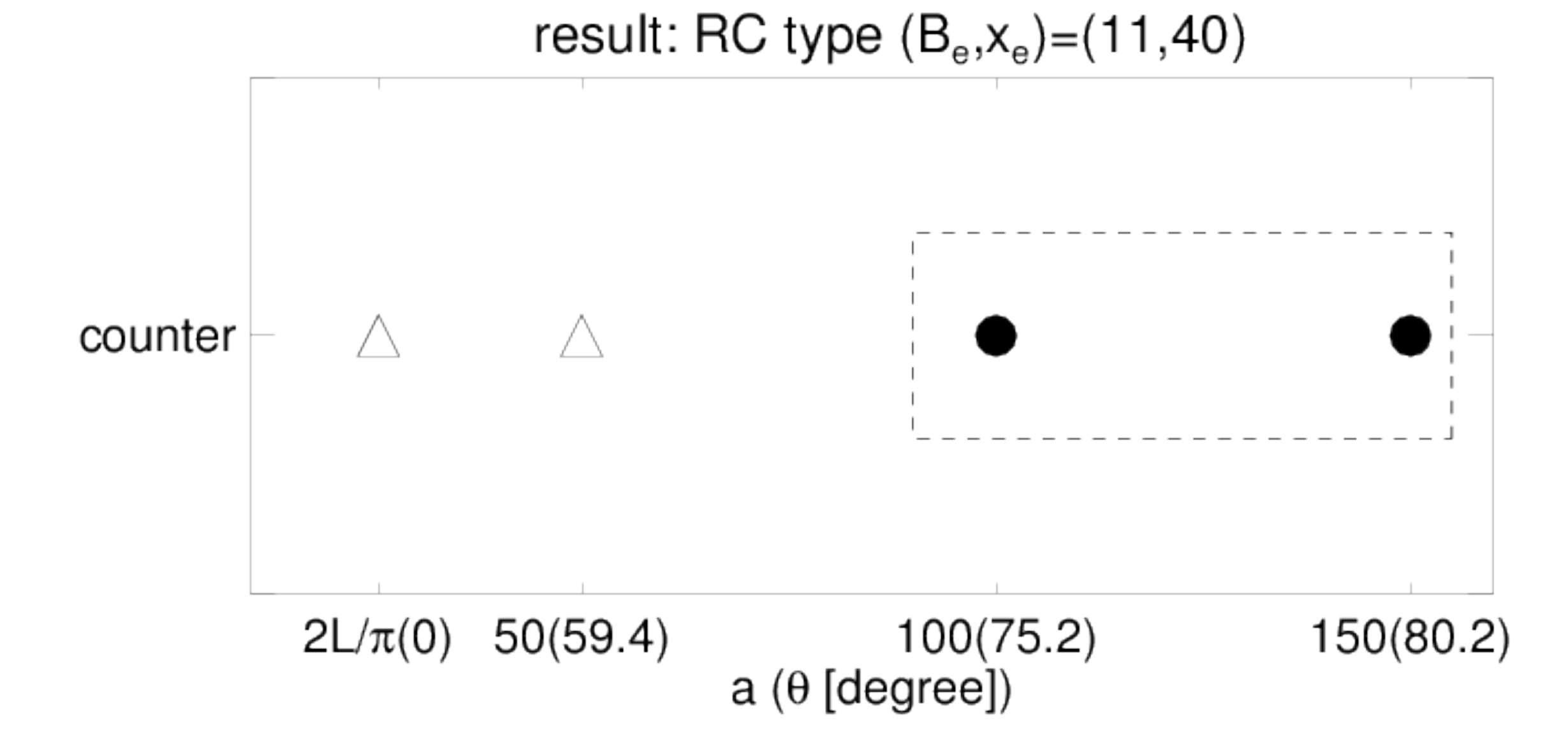}
    \end{center}
  \end{minipage}
\end{tabular}
  \caption{\small{
      Results of parameter survey on  magnetic scale height 
      of the arcade field $a$. Each value 
      of $a$ corresponds to the value of $\theta $ in brackets.
      Note that the arcades are potential fields in the cases of $a=2L/\pi =25.5$. 
      Circles are `erupted', triangles are `formed', and crosses are `unformed' cases.   
      The upper panel shows the results of  CA-type eruptions and
      the lower panel shows the results of RC-type eruptions. 
      Note that RC-type eruptions occur only in  counter-arranged arcades.}}
  \label{resa}  
\end{figure}

In our two-dimensional simulations, the effects of the torus and kink instabilities
(\citealp{2004A&A...413L..23K}, \citealp{2005ApJ...630L..97T},
\citealp{2006PhRvL..96y5002K}, \citealp{2007ApJ...668.1232F}) are absent.
The criteria of the torus instability is $-d \ln{B_{\tani{ext}}} / d \ln{R} \ge n_{\tani{cr}}$,
where $B_{\tani{ext}}$ is the external magnetic field 
strength, $R$ is major radius of toroidal flux rope,  $n_{\tani{cr}}$ is decay index.
For the two-dimensional case of an infinite long, straight flux rope, 
the criteria of eruptions based on the loss of equilibrium is expressed 
by the same manner with $n_{\tani{cr}}=1$
\citep{vanTendKuperus1978SoPh}.
As seen in Fig.\ref{resa}, eruptions are likely to occur when the magnetic scale height $a$ is larger.
This dependence on $a$ will be opposite if the loss of equilibrium or the torus instability is effective.
For the arcade field used in our simulations, the critical condition of the loss of equilibrium of a
plasmoid is
\begin{equation}
\frac{y}{a} \ge n_{\tani{cr}}=1.  \label{ctorus}
\end{equation}
Eq.(\ref{ctorus}) shows that eruptions are more likely to be triggered 
when the magnetic scale height $a$ is smaller (note that the situation is same as when the shear angle $\theta $ is smaller).
The critical height in our simulations is around $y \approx a$.
In the all formed cases (triangles in Fig.\ref{resa}), the formed plasmoids exceed the critical height, however, they do not result in full eruptions.
In the present two-dimensional simulations, we can not find the clear evidence for the occurrence of the loss of equilibrium. It is necessary to confirm the dependence on $a$ in three-dimensional simulations.
If the coupling with the instabilities (not only the torus instability but also the kink instability) is achieved , 
the plasmoid will be accelerated more and result in the faster eruptions such as CME.  

\section{SUMMARY}
The mechanisms of eruptions associated with interaction between 
 newly emerging flux and  coronal arcade fields were investigated 
by 2.5-dimensional MHD simulations. 
We found that two possible eruptive mechanisms exist for the configuration studied here,
and we determined their parametric dependence on the 
magnetic scale height of the arcade field,
the amount of emerging flux, and the location of emerging flux.
The CA-type mechanism, which occurs if emerging flux compresses the
field in the lower part of an arcade, is dominant 
when the location of emerging flux surrounds the edges of arcades.
The condition to trigger the CA-type eruption is represented as formula (\ref{econd}),
namely, the formation and eruption of a plasmoid occurs 
when the location of emerging flux is close to the edge of the arcade field
and the amount of flux introduced into the corona is large enough.
The eruptions are also more likely to occur when the magnetic scale height is large
(this is the same situation as when the arcade is strongly sheared).
The RC-type mechanism, which occurs if multiple reconnections occur in the
current sheet between the emerging flux and the counter-arranged arcade,  
is dominant when emerging flux surrounds the PIL. 
The parameter dependence shows that such eruptions are
more likely to occur when a small amount of magnetic flux is introduced and when the arcade is 
strongly sheared.
The eruptive mechanisms we described here appear to work
in fully three-dimensional configurations as well. For example,
the simulation by \citet{Notoya2007ASPC} suggests the occurrence 
of a CA-type mechanism, while the eruptive mechanism described in
\citet{Roussev2012Nat} corresponds to our RC-type cases.
The kink or torus instability can be expected to support both types of eruptions in three-dimensional cases.

\acknowledgments
We are grateful to the referee for his/her constructive suggestions and tolerant encouragement. 
Numerical computations were carried out on 
general-purpose PC Farm and Cray XC30 at the Center for Computational Astrophysics, 
CfCA, of National Astronomical Observatory of Japan. 
We have greatly benefited from the proofreading/editing assistance from 
the GCOE program.



\clearpage

\end{document}